\documentstyle[epsf]{l-aa}
\begin{document}

\def\clipfig#1{\def\lbracket{[}\def\testit{#1}%
    \ifx\testit\lbracket\let\next=\optclipfig\else\let\next=\stdclipfig\fi%
    \next{#1}}
%
\newcommand {\hclipfig} [7] {\clipfig[#7]{#1}{#2}{#3}{#4}{#5}{#6}}
%
\def\usemodepsfig {\global\def\cfmode{x}\typeout{*** set clipfig to PSFIG mode ***}}
\def\usemodeepsf  {\global\def\cfmode{}\typeout{*** set clipfig to EPSF mode ***}}
\def\useunitmm    {\global\def\cfunit{x}\typeout{*** set clipfig to use mm as unit ***}}
\def\useunitcm    {\global\def\cfunit{}\typeout{*** set clipfig to use cm as unit ***}}
\def\clipfigsettings {\ifx\cfmode\empty\def\ccfmode{EPSF }\else\def\ccfmode{PSFIG }\fi%
    \ifx\cfunit\empty\def\ccfunit{cm }\else\def\ccfunit{mm }\fi%
    \typeout{*** current clipfig settings: \ccfmode mode, using \ccfunit as unit ***}}
%
%
%
%
\def\stdclipfig#1#2#3#4#5#6{\ifx\cfmode\empty%
    \let\next=\eclipfig\else\let\next=\pclipfig\fi%
    \next{#1}{#2}{#3}{#4}{#5}{#6}}
\def\optclipfig#1#2]#3#4#5#6#7#8{\ifx\cfmode\empty%
    \let\next=\ehclipfig\else\let\next=\phclipfig\fi%
    \next{#3}{#4}{#5}{#6}{#7}{#8}{#2}}
%
%
%
\newcommand {\pclipfig}[6] {\ifx\cfunit\empty%
        \psfig{figure=#1.ps,width=#2cm,bbllx=#3cm,bblly=#4cm,bburx=#5cm,%
           bbury=#6cm,clip=}\else%
        \psfig{figure=#1.ps,width=#2mm,bbllx=#3mm,bblly=#4mm,bburx=#5mm,%
           bbury=#6mm,clip=}\fi}
\newcommand {\phclipfig}[7] {\ifx\cfunit\empty%
        \hspace{#7cm}\psfig{figure=#1.ps,width=#2cm,bbllx=#3cm,bblly=#4cm,%
           bburx=#5cm,bbury=#6cm,clip=}\else%
        \hspace{#7mm}\psfig{figure=#1.ps,width=#2mm,bbllx=#3mm,bblly=#4mm,%
           bburx=#5mm,bbury=#6mm,clip=}\fi}
%
%
%
\newcommand {\eclipfig}[6]{%
  \ifx\cfunit\empty\epsfxsize=#2cm\else\epsfxsize=#2mm\fi%
  \epsfclipon\epsfverbosetrue%
  \cfcmtopspts{#3}\cfllxi=\cftempi\cfllxf=\cftempf%
  \cfcmtopspts{#4}\cfllyi=\cftempi\cfllyf=\cftempf%
  \cfcmtopspts{#5}\cfurxi=\cftempi\cfurxf=\cftempf%
  \cfcmtopspts{#6}\cfuryi=\cftempi\cfuryf=\cftempf%
  \def\cfstra{\number\cfllxi.\number\cfllxf}%
  \def\cfstrb{\number\cfllyi.\number\cfllyf}%
  \def\cfstrc{\number\cfurxi.\number\cfurxf}%
  \def\cfstrd{\number\cfuryi.\number\cfuryf}%
  \hbox{\epsfbox[{\cfstra} {\cfstrb} {\cfstrc} {\cfstrd}]{#1.ps}}}
\newcommand {\ehclipfig}[7]{%
  \ifx\cfunit\empty\epsfxsize=#2cm\else\epsfxsize=#2mm\fi%
  \epsfclipon\epsfverbosetrue%
  \cfcmtopspts{#3}\cfllxi=\cftempi\cfllxf=\cftempf%
  \cfcmtopspts{#4}\cfllyi=\cftempi\cfllyf=\cftempf%
  \cfcmtopspts{#5}\cfurxi=\cftempi\cfurxf=\cftempf%
  \cfcmtopspts{#6}\cfuryi=\cftempi\cfuryf=\cftempf%
  \def\cfstra{\number\cfllxi.\number\cfllxf}%
  \def\cfstrb{\number\cfllyi.\number\cfllyf}%
  \def\cfstrc{\number\cfurxi.\number\cfurxf}%
  \def\cfstrd{\number\cfuryi.\number\cfuryf}%
  \ifx\cfunit\empty\hspace{#7cm}\else\hspace{#7mm}\fi%
  \hbox{\epsfbox[{\cfstra} {\cfstrb} {\cfstrc} {\cfstrd}]{#1.ps}}%
  \vspace{-1mm}}
%
%
%
\newdimen\cfllxi \newdimen\cfllyi  \newdimen\cfurxi  \newdimen\cfuryi
\newdimen\cfllxf \newdimen\cfllyf  \newdimen\cfurxf  \newdimen\cfuryf
\newdimen\cftemp \newdimen\cftempi \newdimen\cftempf
\newdimen\cfpspoint \cfpspoint=1bp
%
%
%
\newcommand{\cfcmtopspts}[1]{\ifx\cfunit\empty%
  \cftemp=#1cm\else\cftemp=#1mm\fi%
  \multiply\cftemp10\divide\cftemp\cfpspoint%
  \cftempf=\cftemp\divide\cftemp10\cftempi=\cftemp\multiply\cftemp10%
  \advance\cftempf-\cftemp}
%
%
\def\cfmode{}\def\cfunit{}\clipfigsettings
%

\useunitmm


\newcommand{\inft}{$\infty$}
\newcommand{\vlv}{$\nu L_{\rm V}$}
\newcommand{\lv}{$L_{\rm V}$}
\newcommand{\lx}{$L_{\rm x}$}
\newcommand{\lsoft}{$L_{\rm 250eV}$}
\newcommand{\lhard}{$L_{\rm 1keV}$}
\newcommand{\vlsoft}{$\nu L_{\rm 250eV}$}
\newcommand{\vlhard}{$\nu L_{\rm 1keV}$}
\newcommand{\vlir}{$\nu L_{60\mu}$}
\newcommand{\ax}{$\alpha_{\rm x}$}
\newcommand{\aopt}{$\alpha_{\rm opt}$}
\newcommand{\aoxs}{$\alpha_{\rm ox-soft}$}
\newcommand{\aoxh}{$\alpha_{\rm ox-hard}$}
\newcommand{\airhard}{$\alpha_{\rm 60\mu-hard}$}
\newcommand{\aoxsoft}{$\alpha_{\rm ox-soft}$}
\newcommand{\aio}{$\alpha_{\rm io}$}
\newcommand{\aixs}{$\alpha_{\rm ixs}$}
\newcommand{\aixh}{$\alpha_{\rm ixh}$}
\newcommand{\hb}{H$\beta$}
\newcommand{\nh}{$N_{\rm H}$}
\newcommand{\nhgal}{$N_{\rm H,gal}$}
\newcommand{\nhfit}{$N_{\rm H,fit}$}
\newcommand{\ale}{$\alpha_{\rm E}$}
\newcommand{\cts}{$\rm {cts\,s}^{-1}$}
\newcommand{\pl}{$\pm$}
\newcommand{\kev}{\rm keV}
\newcommand{\rb}[1]{\raisebox{1.5ex}[-1.5ex]{#1}}
\newcommand{\ten}[2]{#1\cdot 10^{#2}}
\newcommand{\msun}{$M_{\odot}$}
\newcommand{\dM}{$\dot{M}$}
\newcommand{\dMM}{$\dot{M}/M$}
\newcommand{\dMedd}{$\dot{M}_{\rm Edd}$}
\newcommand{\dnh}{$\Delta N_{\rm H}$}

\def\lesssim{\mathrel{\hbox{\rlap{\hbox{\lower4pt\hbox{$\sim$}}}\hbox{$<$}}}}
\def\gtrsim{\mathrel{\hbox{\rlap{\hbox{\lower4pt\hbox{$\sim$}}}\hbox{$>$}}}}

\thesaurus{05(02.01.1, 11.01.2, 11.14.1, 11.17.3, 11.19.1)}
\title{New bright soft X-ray selected ROSAT AGN: }
\subtitle{I. Infrared--to--X-ray spectral energy distributions
\thanks{Based in part on observations at the European
Southern Observatory La Silla (Chile) with
the 2.2m telescope of the Max-Planck-Society during MPG and ESO time,
and the ESO 1.5m telescope
}}
\author{D. Grupe\inst{1, }
\thanks{Present
address MPI f\"ur extraterestrische Physik, Garching, FRG; 
guest observer: McDonald Observatory, U of Texas at Austin.
\newline
{}$^{\dagger}$\hspace{3mm} Deceased December 23, 1996.
}
\and K. Beuermann\inst{1,2}
\and H.-C. Thomas\inst{3}
\and K. Mannheim\inst{1}
\and H.H. Fink\inst{2}$^{,\dagger}$
}
\offprints{grupe@usw052.dnet.gwdg.de}
\institute{Universit\"ats-Sternwarte, Geismarlandstr. 11, D-37083 G\"ottingen,
FRG
\and MPI f\"ur extraterrestrische Physik, Giessenbachstr. 6, D-85748 Garching,
FRG
\and MPI f\"ur Astrophysik, Karl-Schwarzschildstr. 1, D-85748 Garching, FRG
}
\date{Received September 19, 1996 / Accepted September 9, 1997}
\maketitle
\markboth{D. Grupe et al.: New Bright Soft X-Ray Selected ROSAT AGN}{ }


\begin{abstract}

We present results of an infrared-to-X-ray study of 76 bright soft X-ray
selected Seyfert galaxies discovered in the ROSAT All-Sky Survey. These objects 
are characterized by steep X-ray spectra in the 0.2-2.0~keV bandpass with power
law energy spectral indices in the range of 1.3 to 8 and a lack of internal
absorption by neutral hydrogen.   Our sample selection based on hardness ratio
yields a mean slope of ${\alpha_{\rm X}} = 2.1\pm 0.1$ ($F_\nu\propto
\nu^{-\alpha}$),  steeper than in any other known AGN population.  At optical
wavelengths, the soft AGN have significantly bluer spectra than a comparison
sample of AGN with a canonical, harder X-ray spectrum, whereas the slope between
5500~\AA~ and 1~keV is the same.  This is consistent with a more pronounced Big
Blue Bump emission component in the soft X-ray selected AGN. The blueness of the
optical spectra increases with the softness of the X-ray spectra {\it and} with
the luminosity, saturating at an approximate $F_\nu\propto \nu^{+0.3}$ spectrum.
Such properties are expected if most of the
Big Blue Bump emission originates in a (comptonized) accretion disk and \dMM~ is
higher than in AGN with a hard X-ray spectrum. 
\keywords{accretion, accretion disks -- galaxies:  active -- galaxies:  nuclei
-- quasars:  general -- galaxies:  Seyfert} 
\end{abstract}

\section{Introduction}

The ROSAT All Sky Survey (RASS) has led to the discovery of a large number of
X-ray sources (Voges et al.~1996, Voges 1997). Even among the brightest of these
a sizable fraction coincides with objects not previously recognized as X-ray
emitters. Of particular interest are those with very soft X-ray spectra because
they are likely not to have been adequately covered by previous X-ray missions.
Soft X-rays probe the high-energy accretion disk continuum and they are
important for the photoionization of the nuclear gas in AGN. In this paper we
discuss the infrared-to-X-ray continua of a high galactic latitude sample of the
brightest soft X-ray selected AGN newly discovered in the RASS. This sample of
76 AGN is drawn from an optically identified sample of bright soft RASS sources
(Thomas et al. in prep., Beuermann et al. in prep.). 

In the $2-20\,\kev$ energy range the X-ray spectra of AGN are characterized by a
hard power-law with an energy spectral slope \ax\,$\sim 1.0$, perhaps lowered by
a reflection component to \ax\,$\simeq 0.7$. Throughout this paper, spectral
indices $\alpha$ are energy indices defined by $F_{\rm
\nu}~\propto~\nu^{-\alpha}$. Several recent studies demonstrated that the
spectral index increases at photon energies below $\sim 1\,\kev$ (e.g. Walter
\& Fink 1993, Puchnarewicz et al. 1995a, Boller et al. 1996), indicating what
appears to be a {\it soft X-ray excess} over an underlying hard power law
component. In spite of an extensive debate there is no generally accepted
explanation of the observed steepening of the X-ray spectra. 

The presence of a soft excess was first noted by Arnaud et al.  (1985) in the
EXOSAT data of the Seyfert 1 galaxy Mkn 841.  In the EINSTEIN era, C\'ordova et
al. (1992) and Puchnarewicz et al.  (1992) found that some serendipitously
discovered AGN possessed ``ultrasoft'' X-ray spectra.  Using RASS data, Walter
\& Fink (1993) showed that in a sample of 58 Sy1 galaxies the soft excess is
part of a {\it Big Blue Bump} (henceforth BBB) which extends from the UV
spectral range to soft X-ray energies.  Because of the rapid spectral
variability seen in many of these systems (e.g.  Boller et al.  1993, 1996), the
soft X-rays must originate from the innermost regions of the AGN.  A popular
interpretation of the BBB is thermal emission from an accretion disk (Shields
1978, Malkan \& Sargent 1982, Malkan 1983, Band \& Malkan, 1989), probably
modified by comptonization which shifts some of the photons into the soft X-ray
regime (Czerny \& Elvis 1987, Laor \& Netzer 1989, Ross et al.  1992, Mannheim
et al. 1995).  Taking a different approach, Ferland \& Rees (1988) and Guilbert
\& Rees (1988) suggested that hard X-ray emission is reprocessed into the
optical-to-soft X-ray range by cold dense clouds, while Barvainis (1993)
proposed that free-free emission from a high-temperature optically thin gas can
account for the spectral shape of the BBB.  A steep soft X-ray spectrum could
also be produced by a warm absorber which preferentially absorbs X-rays at
intermediate energies of $\sim 0.5 - 3\,\kev$, while being transparent at lower
energies (George et al.~1995, Komossa \& Fink 1997). 

\begin{figure}[t]
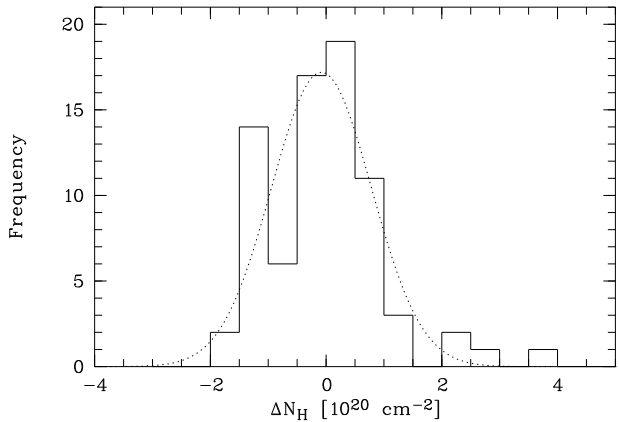

\clipfig{distr_nh2}{85}{10}{10}{280}{195}
\caption[ ]{\label{distr_nh}
Distribution of the difference \nh-\nhgal~ for our soft sample (solid
histogram). The dotted line shows a Gaussian fit to this histogram with the
center at --0.03 and $\sigma$ = 0.87.
}
\end{figure}

In this paper we present new data on ROSAT-discovered soft X-ray selected AGN
from which we expect further clues on the origin of the soft X-rays and the
nature of the BBB. In Sect.~2 we describe the X-ray observations and define the
sample. The analysis of the X-ray data and the results of a correlation study
between X-ray and optical continuum properties are presented in Sect.~3 and
discussed in Sect.~4. All luminosities are calculated with $H_0=75\,{\rm
km\,s^{-1}Mpc^{-1}}$ and $q_0={1\over 2}$. In a separate paper, we present the
optical data and discuss correlations between optical and X-ray continuum
properties and optical  emission line properties (Grupe et al. in prep., 
hereafter Paper II). 

\section{Sample Definition} 

The 76 sources presented in this paper were chosen from RASS data, observed with
the Position Sensitive Proportional Counter (PSPC, Pfeffermann et al. 1986). Our
optical identification program (Thomas et al. in prep., Beuermann et al. in 
prep.) focused on bright soft high galactic latitude sources fulfilling the 
criteria: 
\begin{itemize}
\item galactic latitude $|\,b\,| > 20^{\circ}$,
\item PSPC count rate (CR) during the RASS $\gtrsim$ 0.5\,\cts, 
\item hardness ratio HR1 + error (HR1)$ < 0$, i.e. effectively more
than 50\% of the counts in the Carbon band ($E < 0.28\,\kev$).
\end{itemize}
The objects were chosen to be at high galactic latitudes to avoid cold
absorption in the galactic plane. The high count rate was chosen for two
reasons: first to get reasonable signal-to-noise ratios, and second to get
better positions for reliable identifications. The hardness ratio is defined as
HR1 = (HARD-SOFT)/(HARD+SOFT) where SOFT and HARD are the count rates in the
energy bands $0.1-0.4$ and $0.4-2.4\,\kev$, respectively. HR1 is equivalent to
an X-ray ``colour''. 

All PSPC count rates quoted in this paper are from the RASS Bright Source
Catalogue (RASS BSC) (Voges et al. 1997). The original source selection for our
identification program was based on a limit of 0.5 \cts~ in preliminary count
rates which differ from those which finally entered the RASS BSC. As a
consequence, our sample is not complete in terms of a flux limit. 

The present sample includes 53 AGN which were previously unknown as X-ray as
well a optical objects; it is supplemented by 23 objects which were known to be
AGN in that an entry was found in the NASA/IPAC Extragalactic Database, but for
which no optical spectra or only low-resolution spectra were available in the
literature. Combined, we have a sample of 76 soft X-ray emitting AGN which were
not previously known as X-ray sources and which had largely unknown optical
spectroscopic properties. For most sources, we have taken flux-calibrated
low-resolution identification spectra which cover the range from typically $3800
- 10000$\AA~ and, for the whole sample, we have collected medium-resolution
($\sim5$\AA~ FWHM) optical spectroscopy, using telescopes at La Silla, the
Observatoire de Haute-Provence, Calar Alto, and McDonald Observatory. In this
paper, we make use of the low-resolution spectra which define the shape of the
optical continuum; the emission-line properties will be discussed in Paper II.

\begin{figure}[t]
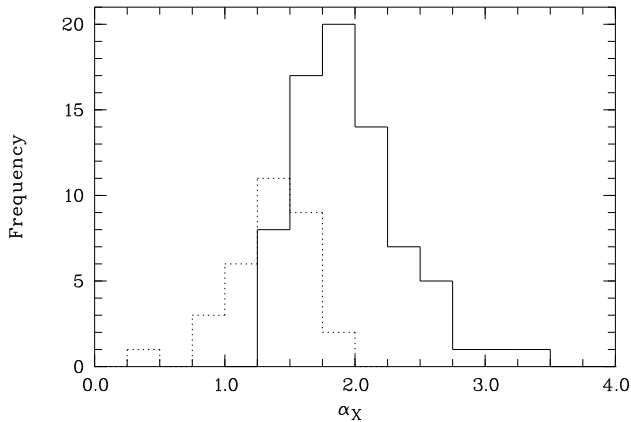

\clipfig{distr_ax}{85}{10}{10}{280}{195}
\caption[ ]{\label{distr_ax}
Distributions of the single power law slopes \ax~ of the unabsorbed 
X-ray spectra for the soft (solid) and
hard sample (dotted). RX J0134 (\ax=6.7), RX J0136 (4.9), and WPVS007 
(8.0) are off the plot.
}
\end{figure}

In order to set the properties of the soft X-ray selected sample of AGN into
perspective, we defined a comparison sample of hard X-ray selected sources
(Grupe 1996)\footnote{ Fairall 009, Mkn 590, Mkn 1044, NGC 985, ESO 198-G24, 3C
120, Mkn 1095, Mkn 79, PG 0804+761, Mkn 704, Mkn 705, NGC 3783, Mkn 766, Mkn
205, ESO 383-G35, IC 4329A, Mkn 279, Mkn 1383, Mkn 876, Mkn 506, EXO 1821+64,
ESO 141-G55, Mkn 509, Mkn 1513, NGC 7213, NGC 7469, Mkn 926 }. All of these have
PSPC count rates $>1$\,\cts~ and hardness ratios HR1$>0$ during the RASS. X-ray
data of these sources were derived from the public ROSAT data archive at MPE
Garching and optical data were taken from the literature. Because we included
only well-known sources our comparison sample is also selected by optical
criteria. Thus we miss any bright ROSAT-discovered hard X-ray AGN and might
exclude AGN with high optical extinction. 

\footnotesize
\begin{table*}
\caption[ ]{\label{parapl} Parameters of the 76 AGN of the soft X-ray sample.
The columns denote (1) running number, (2) the name of the object (full names
are given in Table \ref{lum-list}), (3) the redshift, (4) the visual continuum
magnitude $V$ (with estimated error $\pm0.2$ mag), (5) the optical spectral
index  \aopt~ (with estimated error $\pm0.4$), (6) the exposure time in s, (7)
the RASS PSPC count rate CR in \cts, (8) the hardness ratio HR1, (9) the adopted
atomic hydrogen column density in units of $10^{20}$\rm cm$^{-2}$ (see
footnote$^{3)}$), (10) the logarithm of the rest-frame $0.2-2.0\,\kev$ energy
flux in W\,m$^{-2}$ for a power law spectrum with, (11) X-ray spectral index \ax
(with errors from the fit, see EXSAS user's guide), (12) $\chi^2/\nu$ for the
fit, and (13) \dnh = \nhfit~$-$\nhgal~ in the same units as (9) with errors
from the X-ray spectral fit.} 
\begin{flushleft}
\begin{tabular}{rlccr@{}lrlllclcc}
\hline \noalign{\smallskip}
(1) & (2) & (3) & (4) & (5) & & (6) & (7) & (8) & (9) & (10) & (11) & (12) &
(13)  \\
No & Name$^{1)}$ & $z$ & $V$ & \aopt && $T_{\rm exp}$ & CR & HR1 & \nh
& log\,$F_{\rm x}$ & \ax & $\chi^2/\nu$ & \dnh \\
\noalign{\smallskip}\hline\noalign{\smallskip}
1 & RXJ0022-34 & 0.219 & 16.1 & 0.0& & 471 & 0.47\pl0.05 
& --0.33\pl0.07 & 1.39 & --14.4 & 1.6\pl0.2 & 12/13 & $-$0.4\pl1.3 \\
2 & ESO242-G8 & 0.059 & 16.1&  1.4& & 345 & 0.50\pl0.04  
& --0.41\pl0.10  & 1.48 & --14.2 & 1.6\pl0.2 & 14/12 & $-$0.4\pl1.2 \\
3 & WPVS 007 & 0.029 & 14.8 & 0.6& & 268 & 0.98\pl0.08  & --0.94\pl0.06  & 2.82 
& --14.2 & 8.0\pl2.0 & 10/13 & $-$1.4\pl2.2 \\
4 & RXJ0057--22 & 0.062 & 14.5 & 0.4& & 509 & 2.72\pl0.10  
& --0.50\pl0.03   & 1.48 & --13.7 & 1.9\pl0.1 & 23/29 & +0.0\pl0.5 \\
5 & QSO0056--36 & 0.165 & 15.1 & --0.3& & 704 & 0.65\pl0.03  
& --0.24\pl0.07  & 1.94 & --14.1 & 1.7\pl0.2 & 10/9 & $-$0.2\pl1.1 \\
6 & RXJ0100--51 & 0.062 & 15.4 & 1.0& & 306 & 1.04\pl0.08  
& --0.15\pl0.06  & 2.42 & --13.9 & 1.7\pl0.2 & 9/18 & $-$0.2\pl1.5 \\
7 & MS0117--28 & 0.349 & 16.0 & 0.0& & 721 & 0.32\pl0.02  
& --0.56\pl0.08  & 1.65 & --14.3 & 2.4\pl0.3 & 14/17 & $-$0.4\pl1.0 \\
8 & IRAS01267 & 0.093 & 15.4 & 0.4& & 251 & 0.96\pl0.08  
& --0.40\pl0.07  & 1.28 & --14.2 & 1.6\pl0.2 & 22/15 & +0.1\pl1.4 \\
9 & RXJ0134--42$^{2)}$  & 0.237 & 16.0 & --0.1& & 574 & 0.21\pl0.05  
& --0.84\pl0.05  & 1.59 & --14.9 & 6.7\pl2.6 & 2.2/6 & $-$1.3\pl1.1 \\
10 & RXJ0136--35 & 0.289 & 18.0 & 0.9& & 539 & 0.50\pl0.05   & --0.72\pl0.05 
 &  $5.60^{3)}$ & --13.3 & 4.9\pl0.5$^{3)}$ & 14/16 & +3.8\pl4.0 \\
11 & RXJ0148--27 & 0.121 & 15.5 & 0.6& & 415 & 2.42\pl0.10  
& --0.53\pl0.03  & 1.50 & --13.8 & 2.0\pl0.1 & 26/20 & +0.7\pl0.8 \\
12 & RXJ0152--23 & 0.113 & 15.6 & 0.5& & 702 & 1.06\pl0.07  
& --0.47\pl0.04  & 1.10 & --14.2 & 1.6\pl0.1 & 11/15 & +0.3\pl0.7 \\
13 & RXJ0204--51 & 0.151 & 16.6 & 0.7& & 489 & 0.43\pl0.06  
& --0.22\pl0.08  & 3.76 & --14.1 & 2.3\pl0.3 & 11/12 & $-$1.5\pl2.0 \\
14 & RXJ0228--40 & 0.494 & 15.2 & 0.0& & 748 & 0.57\pl0.06  
& --0.55\pl0.05  & 1.78 & --14.2 & 2.1\pl0.2 & 9.4/9 & +0.0\pl1.2 \\
15 & RXJ0319--26 & 0.079 & 15.9 & 1.3 & & 263 & 0.52\pl0.06  
& --0.41\pl0.09  & 1.32 & --14.5  & 2.0\pl0.3 & 8.9/8 & $-$0.7\pl1.2 \\
16 & RXJ0323--49 & 0.071 & 16.5 & 1.5& & 845 & 1.48\pl0.06  
& --0.44\pl0.03  & 1.72 & --13.9  & 2.0\pl0.1 & 29/26 & +0.3\pl0.7 \\
17 & ESO301--G13 & 0.064 & 15.5 & 1.3& & 898 & 0.81\pl0.03  
&--0.37\pl0.05  & 2.19 & --14.0  & 2.1\pl0.1 & 19/14 & $-$0.4\pl0.9 \\
18 & VCV0331--37 & 0.064 & 16.3 & 1.1& & 353 & 0.59\pl0.04  
& --0.34\pl0.09  & 1.63 & --14.3 & 1.6\pl0.3 & 11/9 & +0.4\pl2.1 \\
19 & RXJ0349--47$^{2)}$  & 0.299 & 16.8 & 1.9& & 449 & 0.48\pl0.06 
& --0.69\pl0.06  & 1.44 & --14.5 & 2.9\pl0.5 & 10/12 & $-$0.2\pl1.1 \\
20 & Fairall 1116 & 0.059 & 15.2 & 0.9& & 222 & 1.49\pl0.09  
& --0.29\pl0.08  & 3.84 & --13.5  & 2.4\pl0.2 & 21/20 & $-$1.1\pl1.8 \\
21 & RXJ0412--47 & 0.132 & 15.9 & 1.0& & 163 & 0.78\pl0.10  
& --0.45\pl0.09  & 1.42 & --14.3 & 1.8\pl0.3 & 3.7/7 & $-$0.7\pl1.3 \\
22 & RXJ0426--57 & 0.104 & 14.1 & 0.0& & 797 & 3.63\pl0.09  
& --0.48\pl0.02  & 2.25 & --13.4  & 2.2\pl0.1 & 68/30 & $-$1.2\pl0.3 \\
23 & Fairall 303 & 0.040 & 16.2 & 1.1& & 384 & 0.66\pl0.05  
& --0.55\pl0.08  & 0.99 & --14.2 & 1.6\pl0.2 & 12/20 & +0.3\pl0.9 \\
24 & RXJ0435--46 & 0.070 & 17.1 & 1.5& & 652 & 0.37\pl0.05  
& --0.57\pl0.06  & 1.80 & --14.6 & 2.2\pl0.3 & 22/13 & $-$1.3\pl0.8 \\
25 & RXJ0435--36 & 0.141 & 17.1 & 1.8& & 471 & 0.45\pl0.06  
& --0.26\pl0.07  & 1.49 & --14.5 & 1.6\pl0.2 & 12/12 & $-$0.7\pl1.1 \\
26 & RXJ0437--47 & 0.052 & 15.3 & 0.2& & 649 & 1.08\pl0.07  
& --0.50\pl0.04  & 1.69 & --14.1 & 2.0\pl0.2 & 17/14 & $-$1.0\pl0.5 \\
27 & RXJ0438--61 & 0.069 & 15.7 & 0.4& & 2124 & 0.59\pl0.04  
& --0.14\pl0.03  & 2.93 & --14.1 & 1.9\pl0.1 & 26/25 & $-$1.1\pl0.7 \\
28 & RXJ0439--45 & 0.224 & 16.6 & 0.4& & 632 & 0.41\pl0.05  
& --0.77\pl0.04  & 2.05 & --14.4 & 3.3\pl0.7 & 17/13 & $-$1.3\pl0.9 \\
29 & RXJ0454--48$^{2)}$  & 0.363 &  17.7 & 0.4& & 452 & 0.11\pl0.05  
& --0.62\pl0.11  & 1.91 & --14.9 & 2.4\pl0.7 & 0.9/2 & +0.6\pl5.9 \\
30 & RXJ1005+43 & 0.178 & 16.4 & 0.1&$^{4 }$ & 694 & 0.76\pl0.05  
& --0.58\pl0.04  & 1.08 & --14.3 & 1.9\pl0.2 & 15/11 & +0.6\pl0.9 \\
31 & CBS 126 & 0.079 & 15.4 & 0.7& & 853 & 1.22\pl0.04  
& --0.33\pl0.04  & 1.41 & --14.0 & 1.6\pl0.1 & 28/20 & $-$0.0\pl0.6 \\
32 & RXJ1014+46$^{2)}$  & 0.324 & 17.1 & 1.5&$^{5 }$ & 715 & 0.19\pl0.05  
& --0.67\pl0.08  & 0.93 & --15.0 & 2.0\pl0.5 & 5.0/6 & +1.1\pl2.7 \\
33 & RXJ1017+29 & 0.049 & 15.7 & 1.6&$^{4 }$ & 498 & 0.39\pl0.05  
& --0.31\pl0.08  & 2.61 & --14.4 & 2.0\pl0.2 & 7.3/12 & +0.2\pl2.4 \\
34 & Mkn 141 & 0.042 & 15.1 & 1.5& & 975 & 0.43\pl0.02  & --0.38\pl0.07  
& 1.07 & --14.5  & 1.5\pl0.2 & 22/18 & +0.2\pl1.2 \\
35 & Mkn 142 & 0.045 & 15.2 & 1.2& & 1034 & 0.98\pl0.03  
& --0.53\pl0.04  & 1.18 & --14.0  & 1.8\pl0.1 & 31/25 & +0.5\pl0.6 \\
36 & RXJ1050+55 & 0.333 & 16.7 & 1.5&$^{5 }$ & 804 & 0.41\pl0.05  
& --0.75\pl0.04  & 0.63 & --14.7  & 1.9\pl0.3 & 16/16 & +0.4\pl0.9 \\
37 & EXO1055+60 & 0.149 & 16.9 & 1.3&$^{4 }$ & 874 & 0.43\pl0.05  
& --0.70\pl0.04  & 0.60 & --14.8 & 1.9\pl0.3 & 14/19 & $-$0.2\pl0.6 \\
38 & RXJ1117+65 & 0.147 &  16.4 & 1.2& & 897 & 0.59\pl0.05   & --0.66\pl0.04  
& 0.91 & --14.5  & 1.9\pl0.2 & 7.9/10 & +0.8\pl1.0 \\
39 & Ton 1388 & 0.177 & 14.4 & 0.6& & 466 & 0.96\pl0.07  & --0.46\pl0.05  
& 1.22 & --14.2 & 1.7\pl0.2 & 10/9 & +0.2\pl0.9 \\
40 & Mkn 734 & 0.033 & 14.4 & 0.9& & 807 & 0.37\pl0.02  & --0.28\pl0.08  
& 2.64 & --14.3  & 2.0\pl0.2 & 16/14 & $-$1.3\pl1.2 \\
41 & Z1136+34 & 0.033 & 16.0 & 1.6&$^{4 }$ & 591 & 0.42\pl0.03  
& --0.30\pl0.09  & 2.04 & --14.2 & 1.9\pl0.2 & 10/18 & $-$0.4\pl1.2 \\
42 & CSO 109 & 0.059 &  16.3 & 1.4&$^{4 }$ & 863 & 0.38\pl0.02  
& --0.30\pl0.08  & 1.96 & --14.4  & 1.7\pl0.2 & 22/16 & +1.2\pl2.2 \\
43 & RXJ1231+70 & 0.208 & 16.0 & 1.6&$^{5 }$ & 1054 & 1.05\pl0.05  
& --0.14\pl0.03  & 1.67 & --14.0  & 1.4\pl0.1 & 21/23 & +0.2\pl0.7 \\
44 & IC 3599 & 0.021 & 16.5 & 1.8& & 670 & 4.91\pl0.11   & --0.55\pl0.02  
& $3.77^{3)}$ & --13.1 & 3.2\pl0.1$^{3)}$ & 70/42 & +2.7\pl0.8 \\
45 & IRAS1239+33 & 0.044 & 15.1 & 2.4&$^{4 }$ & 529 & 1.08\pl0.06  
& --0.08\pl0.05  & 1.35 & --14.2  & 1.3\pl0.1 & 16/12 & +0.7\pl1.2 \\
46 & RXJ1312+26 & 0.061 & 16.2 & 1.4&$^{4 }$ & 694 & 0.57\pl0.05  
& --0.45\pl0.05  & 1.10 & --14.5 & 1.5\pl0.2 & 17/20 & $-$0.1\pl0.9 \\
47 & RXJ1314+34 & 0.075 & 16.3 & 0.8& & 758 & 0.70\pl0.05  & --0.54\pl0.04  
& 0.99 & --14.5 & 2.0\pl0.2 & 21/11 & +0.7\pl1.0 \\
48 & RXJ1355+56 & 0.122 & 16.5 & 1.7&$^{4 }$ & 842 & 0.71\pl0.06  
& --0.56\pl0.06  & 1.15 & --14.4 & 1.9\pl0.2 & 13/12 & +0.8\pl1.0 \\
49 & RXJ1413+70 & 0.107 & 16.9 & 2.9&$^{4 }$ & 1028 & 0.83\pl0.05  
& --0.16\pl0.05  & 1.93 & --14.1  & 1.5\pl0.1 & 18/17 & $-$0.8\pl0.7 \\
50 & Mkn 684 & 0.046 & 14.7 & 1.2& & 810 & 0.51\pl0.03  & --0.17\pl0.07  
& 1.50 & --14.4 & 1.4\pl0.2 & 20/18 & +0.4\pl1.4 \\
\noalign{\smallskip}\hline\noalign{\smallskip}
\end{tabular}
\end{flushleft}
\end{table*}
\begin{table*}
\begin{flushleft}
\begin{tabular}{rlccr@{}lrlllclcc}
\hline \noalign{\smallskip}
(1) & (2) & (3) & (4) & (5) & & (6) & (7) & (8) & (9) & (10) & (11) & (12) &
(13)  \\
No & Name$^{1)}$ & $z$ & $V$ & \aopt && $T_{\rm exp}$ & CR & HR1 & \nh
& log\,$F_{\rm x}$ & \ax & $\chi^2/\nu$ & \dnh \\
\noalign{\smallskip}\hline\noalign{\smallskip}
51 & Mkn 478 & 0.077 & 14.6 & 0.7& & 926 & 2.98\pl0.06  & --0.65\pl0.02  
& 1.04 & --13.6 & 2.1\pl0.1 & 35/33 & +0.2\pl0.3 \\
52 & RXJ1618+36 & 0.034 & 16.6 & 1.6&$^{4 }$ & 1140 & 0.86\pl0.05  
& --0.36\pl0.03 & 1.26 & --14.2  & 1.5\pl0.1 & 12/19 & +0.0\pl0.6 \\
53 & RXJ1646+39 & 0.100 & 17.1 & 1.3&$^{4 }$ & 4086 & 0.43\pl0.01  
& --0.40\pl0.03 & 1.40 & --14.5  & 1.6\pl0.2 & 22/17 & +0.7\pl1.5 \\
54 & RXJ2144--39 & 0.140 & 18.0 & 1.2& & 496 & 0.43\pl0.05   & --0.42\pl0.07 
&$4.89^{3)}$& --13.8 & 3.4\pl0.3$^{3)}$ & 12/11 & +2.5\pl3.7 \\
55 & RXJ2154--44 & 0.344 & 15.8 &  0.2& & 521 & 0.46\pl0.05  & --0.64\pl0.06  
& 2.07 & --14.3 & 2.5\pl0.4 & 16/12 & $-$1.4\pl0.8 \\
56 & RXJ2213--17 & 0.146 & 17.2 & 1.1& & 373 & 0.49\pl0.06  & --0.57\pl0.07  
& 2.48 & --14.3 & 2.4\pl0.4 & 13/11 & $-$1.8\pl0.9 \\
57 & RXJ2216--44 & 0.136 & 15.8 & 0.2& & 486 & 1.45\pl0.08  
& --0.61\pl0.03 & 2.17 & --13.9 & 2.5\pl0.2 & 24/15 & $-$1.3\pl0.5 \\
58 & RXJ2217--59 & 0.160 & 16.2 & 0.3& & 603 & 0.81\pl0.06  & --0.60\pl0.04  
& 2.58 & --14.0  & 2.7\pl0.2 & 8.9/10 & $-$1.1\pl0.9 \\
59 & RXJ2221--27$^{2)}$  & 0.177 & 17.7 & 0.5& & 417 & 0.29\pl0.06  & 
--0.60\pl0.08
& 1.42 & --14.7 & 2.1\pl0.4 & 5.0/7 & $-$0.8\pl1.2 \\
60 & RXJ2232--41$^{2)}$  & 0.075 & 16.9 & 0.9& & 442 & 0.22\pl0.05 & 
--0.48\pl0.10 
& 1.60 & --14.8  & 1.8\pl0.5 & 8.6/5 & $-$0.4\pl2.1 \\
61 & RXJ2241--44 & 0.545 & 15.8 & 0.3& & 467 & 0.40\pl0.05 & --0.68\pl0.06 
& 1.76 & --14.3  & 2.5\pl0.4 & 14/10 & $-$0.4\pl1.4 \\
62 & RXJ2242--38 & 0.221 & 16.9 & 1.0& & 454 & 0.66\pl0.06 & --0.58\pl0.05 
& 1.18 & --14.4  & 2.2\pl0.3 & 14/15 & +0.7\pl1.4 \\
63 & RXJ2245--46 & 0.201 & 14.8 & 0.8& & 477 & 1.65\pl0.08& --0.64\pl0.03
& 1.95 & --13.8  & 2.5\pl0.2 & 39/14 & $-$1.3\pl0.4 \\
64 & RXJ2248--51 & 0.102 & 15.5 & 1.3& & 503 & 2.62\pl0.13 & --0.54\pl0.03
& 1.27 & --13.8 & 1.9\pl0.1 & 15/25 & $-$0.5\pl0.4 \\
65 & MS2254--37 & 0.039 & 15.0 & 1.9& & 369 & 1.19\pl0.06 & --0.52\pl0.06 
& 1.15 & --14.0 & 1.8\pl0.1 & 33/13 & +0.3\pl0.8 \\
66 & RXJ2258--26 & 0.076 & 16.1 & 1.3& & 145 & 0.69\pl0.09 & --0.11\pl0.12 
& 2.11 & --14.1  & 1.5\pl0.3 & 3.7/6 & +1.3\pl3.8 \\
67 & RXJ2301--55 & 0.140 & 15.4 & 0.1& & 567 & 0.88\pl0.07 & --0.54\pl0.04 
& 1.54 & --14.2 & 2.1\pl0.2 & 6.6/10 & +0.0\pl0.8 \\
68 & RXJ2303--55 & 0.084 & 17.5 & 1.9& & 527 & 0.35\pl0.05 & --0.21\pl0.08
& 1.54 & --14.5 & 1.4\pl0.2 & 13/10 & +0.2\pl1.9 \\
69 & RXJ2304-35 & 0.042 &  16.6 & 1.4& & 205 & 0.56\pl0.07 & --0.37\pl0.10 
& 1.47 & --14.4 & 1.5\pl0.1 & 3.7/6 & $-$0.2\pl2.0 \\
70 & RXJ2304-51 & 0.106 &  17.2 & 1.1& & 418 & 0.38\pl0.06 & --0.40\pl0.09 
& 1.33 & --14.6 & 1.8\pl0.3 & 4.3/9 & $-$0.3\pl1.4 \\
71 & RXJ2317--44$^{2)}$  & 0.134 & 16.8 & 0.3& & 270 & 0.73\pl0.08 &
 --0.74\pl0.05 
& 1.89 & --14.3 & 2.6\pl0.4 & 14/11 & $-$0.9\pl1.1 \\
72 & RXJ2325--32 & 0.216 &  17.0 & 0.8& & 130 & 0.58\pl0.09  & --0.58\pl0.11 
& 1.33 & --14.4 & 2.1\pl0.5 & 5.0/4 & +0.5\pl3.2 \\
73 & RXJ2340--53$^{2)}$  & 0.321 & 17.6 & 0.1& & 231 & 0.29\pl0.06  & 
--0.71\pl0.10 
& 1.81 & --14.5 & 2.1\pl0.6 & 9.4/3 & $-$1.1\pl2.0 \\
74 & MS2340--15 & 0.137 & 15.6 & 0.8 & & 457 & 0.74\pl0.06  & --0.39\pl0.06 
& $4.36^{3)}$ & --13.7 & 3.2\pl0.2$^{3)}$ & 17/16 & +2.2\pl2.5 \\
75 & RXJ2349--33 & 0.135 & 16.6 & 1.3& & 491 & 0.49\pl0.05 & --0.49\pl0.07 
& 1.04 & --14.5 & 1.6\pl0.2 & 7.7/13 & +0.9\pl1.7 \\
76 & RXJ2349-31 & 0.135 & 16.6 & 1.6& & 462 & 0.60\pl0.06 & --0.45\pl0.06 
& 1.23 & --14.4 & 1.6\pl0.2 & 15/16 & $-$0.0\pl1.1 \\
\noalign{\smallskip}\hline\noalign{\smallskip}
\end{tabular}

\noindent {\footnotesize $^{1)}$Exact coordinates of the new sources will be 
given in paper II}\\
\noindent {\footnotesize $^{2)}$ These sources have original 
count rates below the  limit that defined the sample. } \\
\noindent {\footnotesize $^{3)}$\nh\,=\,\nhfit~and \ax~from
\nhfit; for all others \nh\,=\,\nhgal~and
\ax~from \nhgal}\\
\noindent {\footnotesize $^{4)}$ $\alpha_{\rm opt}$ determined from the 1994
McDonald spectra covering 1400 \AA~ only}\\
\noindent {\footnotesize $^{5)}$ Same, but additional uncertainty due
to the larger redshifts (not included in Figs. \ref{lx_aox} and
\ref{alpha_opt-x})}
\end{flushleft}
\end{table*}
\normalsize

\section{Data Analysis}

For all X-ray reduction and analysis tasks we used the EXSAS data analysis
package of the MPE Garching (Zimmermann et al. 1994). The X-ray spectra were
derived from photon event tables covering the area of the sky around individual
sources.  Source counts were collected in a circle of 250 arcsec radius around
the source position, and background counts from a circular region of 750 arcsec
diameter with center offset 825 arcsec from the object position in the scan
direction of the satellite and free of contaminating sources. Although the
spectral fits extend nominally over the energy range $0.1-2.4\,\kev$, the number
of counts below 0.2~keV and above 2.0~keV become very small, and the power law
representation correspondingly uncertain. Unabsorbed fluxes and luminosities
were, therefore, calculated for the restricted energy range of $0.2-2.0\,\kev$. 
Count rates CR and hardness ratios HR1 were derived along with the spectra and
were generally found to be consistent with those given (later) in the RASS BSC
(Voges et al. 1997). For consistency with published values we quote here CR and
HR1 from the BSC. 

Because of the  limited signal-to-noise ratio of the RASS spectra we employed
only single power law fits in the energy range $0.1-2.4\,\kev$ with absorption
by a column density \nh~ of cold material with solar abundances. The hydrogen
column density was either taken equal to the galactic column density
\nhgal~(Dickey \& Lockman 1990) or left as a free parameter, \nhfit. As
explained in Sect. 4.1, \nh~=\nhgal~ is the appropriate choice for most of our
objects. 

The optical spectral index \aopt~is calculated from our low-resolution spectra
(after subtraction of the FeII emission line complex, see Paper II), using the
continuum flux densities at 4400\AA~ and 7000\AA. Comparing the spectral flux
densities with CCD photometry in the BVR bands available for a subset of
objects, we estimate uncertainties of $< 20$\% in the absolute value of the
visual flux and $\sim10$\% in the relative flux calibration between 4400 and
7000\AA~ which translates into an error in \aopt~ of about $\pm 0.4$. 

We have derived the mean overall energy distributions for the objects of the
soft and the hard samples using the mean monochromatic luminosities  $\nu
L_{\nu}$ at selected wavelengths (henceforth abbreviated  `luminosities').  The
IR luminosities for 71 of the sources at 12, 25, 60, and 100 $\mu$m were derived
from the  IRAS survey data using the ADDSCAN procedure of the Infrared
Processing and Analysis Center (IPAC) at Caltech (the missing sources are in
IRAS data gaps). Three-sigma upper limits are quoted if the signal strength is
less than 3 $\sigma$ of the rms fluctuations. Most sources are weak with
exceptional cases exceeding 25\,$\sigma$. The visual luminosities $\nu L_{\rm
V}$ were determined from our optical spectra and CCD photometry. The X-ray
luminosities at 250 eV and at 1 keV were calculated from the power law fits to
the RASS spectra.  All luminosities are in the rest-frame (assuming $\alpha = 1$
in case of the IR data). 

\section{Results}

Table~\ref{parapl} gives the redshift, the visual magnitude $V$, and the power
law index \aopt~ of the optical spectrum,  and X-ray properties of the 76 AGN.
The IR, optical and X-ray luminosities are given in Table~\ref{lum-list} and the
means and medians of the luminosities and spectral indices of the soft and hard
samples are summarized in Table~\ref{mean}. Finally, Table~\ref{rank} lists the
correlations found among the continuum parameters of the soft X-ray sample. 

\subsection{\label{rass} X-ray spectra} 

In columns $6-8$ of Table~\ref{parapl}, we list the exposure time, the mean RASS
PSPC count rate CR and the hardness ratio HR1 from the RASS BSC (Voges et al.
1997). In columns $9-13$, we give the adopted column density of atomic hydrogen,
the logarithm of the $0.2-2.0\,\kev$ rest-frame energy flux, the energy spectral
index \ax~of the best-fit power law, $\chi^2$ for $\nu$ degrees of freedom, and
\dnh~= \nhfit~$-$\nhgal. The uncertainty in log\,$F_{\rm x}$ is determined mainly
by the error in CR while the uncertainties in the monochromatic luminosities
\lsoft~ and \lhard~ depend mainly on the errors in CR and \ax~ (and, of course, 
in  $H_0$).

Figure~\ref{distr_nh} shows the distribution of \dnh~ for the present sample. The
mean values of \nhfit~ and \nhgal~ coincide within the errors, \nhfit =
$\ten{(1.62\pm0.11)}{20}$ $\rm cm^{-2}$ and \nhgal = $\ten{(1.67\pm0.08)}{20}$
$\rm cm^{-2}$, i.e., the mean value of \dnh~ is consistent with zero. The width
of the distribution agrees with that expected from the errors of the spectral
fits as given by the EXSAS program (Zimmermann et al. 1994). We conclude that
there is no evidence for significant internal absorption in the majority of
individual AGN of this sample. This is why we consider \nh~=\nhgal~and the
resulting \ax~as the appropriate choice for most of our soft X-ray selected AGN.
Only four sources have \dnh\,$> \ten{2}{20}$\,cm$^{-2}$ (RXJ0136-35, IC3599,
RXJ2144-39, and MS2340-15) and for these we choose \nh\,=\nhfit~ with \ax~ being
derived from the free fit. Figure~\ref{distr_ax} shows the corresponding
distribution of the derived power law slopes for the soft X-ray sample along
with those for the hard comparison sample. The values for RX J0136-35 and for
the extremely soft transient sources WPVS007 (Grupe et al.~1995a) and RX
J0134-42 (Mannheim et al.~1996) are off the plot. A final caveat is in place,
because we do find some systematic dependence of the derived \ax~ on \dnh, an
effect which is most likely caused by the curvature of the steep X-ray spectra
towards low energies, forming an excess over the power law at the higher
energies. We will show in Sect. 5.2 that \dnh~ can be used as a parameter to
describe the soft X-ray excess. A more detailed analysis would require better
X-ray spectra than obtained from the RASS. 

\begin{figure}[t]
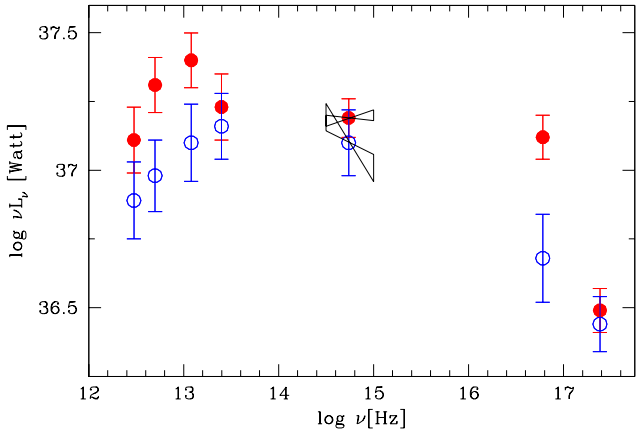

\clipfig{slopes}{87}{10}{90}{195}{215}
\caption[Optical to X-ray slopes Models]{\label{slopes2}
Spectral energy distributions of soft X-ray ({\it solid circles})
and hard X-ray ({\it open circles}) selected AGN, respectively.
Error bars represent the standard errors of the mean.
}
\end{figure}

\begin{figure}[t]
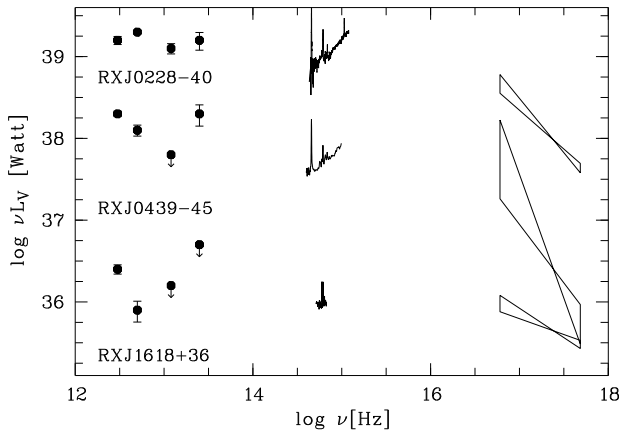
 
\clipfig{sed_example}{87}{7}{0}{277}{195}
\caption[]{\label{sed_example} Examples of (rest-frame) spectral energy
distributions of soft X-ray selected NLSy1 galaxies.  
Arrows indicate 3-$\sigma$
upper limits. } 
\end{figure}

\subsection{\label{mean-lum} Overall energy distributions}

The luminosities of each individual source are given in Table~\ref{lum-list}. The
mean luminosities are listed in Table~\ref{mean} and plotted in
Figure~\ref{slopes2}. The optical reference wavelength is 5500~\AA, the soft and
hard X-ray reference energies are 250~eV and 1~keV, respectively. The mean
IR-luminosities of the soft sample were calculated by a Kaplan-Meier estimator
(see Babu \& Feigelson 1996) which allows to take the upper limits into account.
The visual and hard X-ray luminosities of both samples happen to agree
(Table~\ref{mean} and Fig.~\ref{slopes2}), whereas the soft X-ray and
IR-luminosities of the soft AGN exceed those of the hard comparison sample. 

Also given in Table~\ref{mean} are \aopt~ and \ax~ along with several two-point
spectral indices, connecting different spectral regions. The mean \ax~of our
soft sample (2.11\pl0.10) agrees with that of the optically selected NLSy1
sample of Boller et al. (1996) (weighted mean $\alpha_{\rm x} = 2.13\pm 0.03$).
The mean \ax~of the hard comparison sample (1.35\pl0.06), on the other hand,
agrees reasonably well with the corresponding values for the optically selected
Sy1 sample of Walter \& Fink (1993), $1.50\pm 0.07$ (with standard error
quoted), and the IR selected sample of Rush et al.~(1996), 1.26$\pm$0.11. Laor
et al.~(1997) obtain $\alpha_{\rm x}= 1.63\pm 0.09$ for a sample of optically
selected quasars. 

\begin{figure}[t]
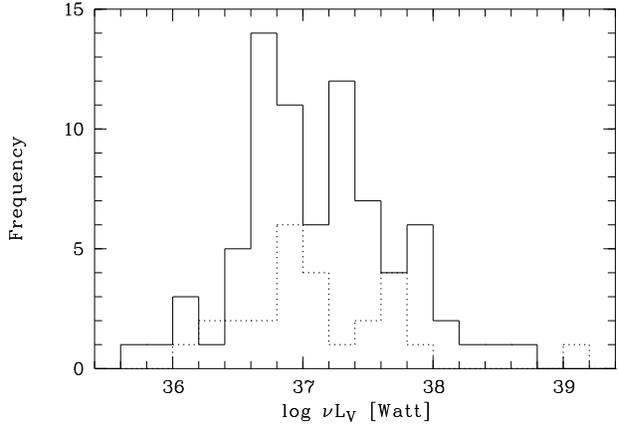
 
\clipfig{distr_lv}{85}{10}{10}{280}{195} 
\caption[ ]{\label{distr_lv} Distributions of the optical monochromatic
rest-frame luminosities $\nu L_V$ at 5500 \AA~ for the soft (solid) and hard
sample (dotted).} 
\end{figure} 

The soft and hard samples have the same mean optical to hard X-ray slopes \aoxh.
They differ in their X-ray spectral indices \ax, optical spectral indices \aopt,
and optical-to-soft X-ray spectral indices \aoxs. The infrared spectral indices
are only marginally different (see Fig. \ref{slopes2}). 

Figure~\ref{sed_example} displays the spectra of three NLSy1 galaxies which
bracket the \ax~ vs.~\lsoft~ correlation shown in Fig.~\ref{lx_aox}. Obviously,
there exists a large diversity of the spectral shapes of individual sources. As
a general trend, however, the BBB becomes more pronounced at higher
luminosities. 

Figure~\ref{distr_lv} displays the distributions of the visual luminosities \vlv~
of the soft and hard sample.  The falloff towards high luminosities represents
the local luminosity function, the falloff towards low luminosities is produced
by our cutoff in PSPC count rate.  As a consequence, our sample is biased in
luminosity and in \ax, by virtue of the hardness ratio cut which removes most
sources with canonical Seyfert X-ray spectra.

\subsection{Correlations}

Correlations between the various monochromatic luminosities and two-point
spectral indices were studied employing the Spearman rank-order test.  The
results are given in Table~\ref{rank}. 

\begin{figure}[t]
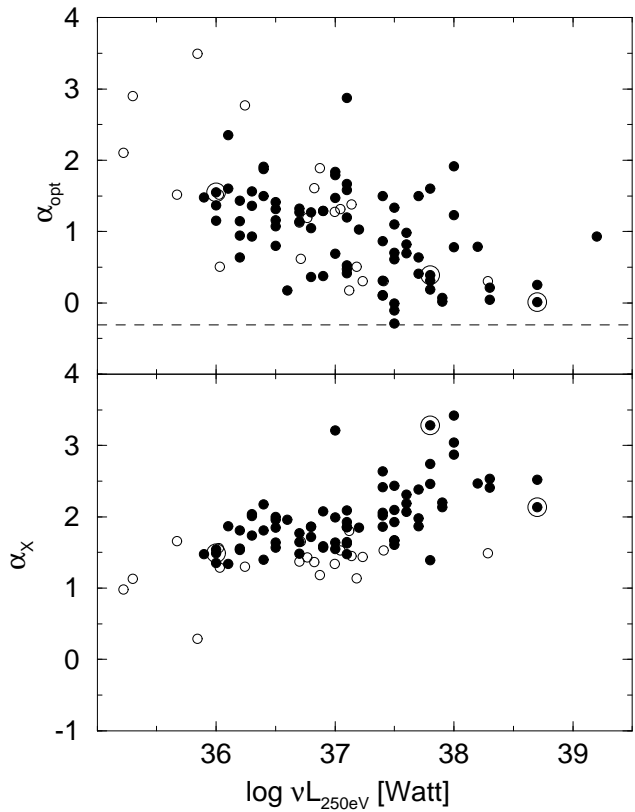

\clipfig{lx_aox}{87}{40}{8}{164}{165}
\caption[ ]{\label{lx_aox} Optical  and X-ray
spectral index  vs.~250~eV luminosity.  The dashed line indicates
the saturation value $\alpha_{\rm opt}$ = $-$0.3 expected for a bare accretion
disk spectrum.  The spectral energy distributions of the sources
corresponding to the encircled points are shown in Fig.~\ref{sed_example}.
The dispersion in $\alpha_{\rm opt}$
is larger than expected from the uncertainties  
($\pm 0.4$, see Sect.~3.1). 
Solid circles denote AGN from the soft, and open circles
from the hard comparison sample.
WPVS007 and RX J0134--42 are off the plot 
(\ax=8.0, log $\nu L_{\rm 250eV}$=36.2; 6.7, 37.5, respectively). 
} 
\end{figure}

Figure~\ref{lx_aox} shows \aopt~and \ax~vs.~\lsoft. The optical index
\aopt~displays a clear variation with  luminosity in the sense that the spectrum
becomes bluer with increasing 250~eV luminosity\footnote{A similar result was
previously reported by Barvainis (1993) for the soft X-ray sample of
Puchnarewicz et al.  (1992) which suffered, however, from problems in the
processing of the EINSTEIN IPC data (see Thompson et al.  1994).}.  The
significance of this anti-correlation is high ($t=-5.3$, see Table~\ref{rank}).
The spectral energy distributions of AGN of low and high luminosity are
considerably different as demonstrated in Fig.~\ref{sed_example}. Since \lv~
correlates with \lsoft~ ($t=14.0$),  the correlations between the spectral
indices and  \lv~  are similarly significant, see Table~\ref{rank}. Furthermore,
for \ax$\simeq 2$, \lx($0.2-2$\,keV) is mainly dependent on the spectral flux at
low energies, hence, \lsoft. The correlations between \ax~and the luminosities
and the anti-correlations between \aopt~ vs. luminosities imply an
anti-correlation between both spectral indices, if the relation is real. 
Spearman's rank order test indeed reveals a convincing anti-correlation at a
significance level of $t=-3.9$ (Table~\ref{rank}). 
We will see below that \dnh~  appears to be related to a soft X-ray excess
showing a weaker correlation with the optical spectral slope ($t=3.1$). 

Although \aopt~ could be influenced by the continuum of the host galaxy,
we see no spectral features indicative of the host
galaxy. The nuclear property \ax~ also shows
a systematic variation with {\lsoft} at the significance level of $t=6.7$. 
Cold, dusty gas might be expected to steepen the optical continuum and 
depress the soft X-rays, producing a correlation between \aopt~ and \ax~ 
in the sense that we observe. However, a neutral hydrogen column of 
$\rm 10^{20}~cm^{-2}$, corresponding to $\rm E_{B-V}\sim$0.02, would 
produce a change of only 0.08 in \aopt, so absorption by cold, dusty gas
is not the cause of the correlation. This is further supported by a lack
of any correlation between \ax~ and the H$\alpha$/H$\beta$ intensity ratio
(Paper II, Grupe et al. 1997). We are aware that a warm dusty
absorber could redden the optical continuum without affecting the Balmer
decrement.  
However, as pointed out in Sect.~4.3, a warm absorber would tend to
{\em steepen} the X-ray spectrum, contrary to what is required to explain the
anti-correlation between \aopt~ and \ax.  A selection
effect is not responsible for the lack of low-luminosity, steep-spectrum X-ray
sources.  Redshift effects are unimportant, since the average redshift is only
$z=0.14$ (Table 2) (e.g.  Schartel et al.  1996). 

\footnotesize
\begin{table*}
\caption{\label{lum-list}
Logarithms of  monochromatic rest-frame luminosities $\nu L_\nu~[W]$ for the
soft X-ray selected AGN listed in Table 1. Missing IR luminosities are due to IRAS
data gaps. The upper limits denote fluxes less than 3-$\sigma$ of the rms
fluctuations. 
}
\begin{flushleft}
\begin{tabular}{rlr@{}lr@{}lr@{}lr@{}llll}
\hline 
\noalign{\smallskip}
\ \ No. & Name && $\nu L_{100\mu}$ && $\nu L_{60\mu}$ && $\nu L_{25\mu}$ && 
$\nu L_{12\mu}$ & $\nu L_{\rm V}$ & $\nu L_{250\rm eV}^{1)}$  & $\nu L_{\rm 1 
keV}^{1)}$  \\
\noalign{\smallskip}\hline\noalign{\smallskip}
1   &  RX J0022.5--3407 & & 
38.7 $^{+.04}_{-.05}$ & &  38.5 $^{+.04}_{-.05}$ & $<$&38.3   & $<$&38.3 	
& 37.7 		& 37.5	 	& 37.1 $^{+0.11}_{-.15}$ \\[0.5ex]
2   & ESO 242-G8 & 
$<$&37.0 & & 37.1$^{+.10}_{-.14}$ &  & 37.1$^{+.07}_{-.08}$   & $<$&37.2
& 36.6 		& 36.5	 	& 36.1$^{+.11}_{-.14}$ \\[0.5ex] 
3   & WPVS 007 &  
& ---    	&& --- 	 && ---     && --- 	
& 36.5 		& 36.2$^{+.15}_{-.24}$	 	& 32.0 \\[0.5ex] 
4   & RX J0057.3--2222 & $<$&36.7 & &36.9$^{+.09}_{-.10}$  
&  &37.8$^{+.03}_{-.03}$   & & 37.3$^{+.12}_{-.16}$	
& 37.3 		& 37.1	 	& 36.6$^{+.05}_{-.06}$ \\[0.5ex] 
5   & QSO0056--363 &
$<$&37.8 & &37.9$^{+.05}_{-.06}$ & $<$&37.9 & $<$&38.2 	
& 37.9 		& 37.5	 	& 37.1$^{+.07}_{-.08}$ \\[0.5ex] 
6   &  RX J0100.4--5113 & &  37.4$^{+.06}_{-.06}$	
& & 37.2$^{+.04}_{-.04}$  &  & 37.4$^{+.03}_{-.04}$   & $<$&37.2 
	& 36.9 		& 36.8	 & 36.4$^{+.08}_{-.10}$ \\[0.5ex] 
7   & MS 0117.2--2837 &
$<$&38.5	& $<$&38.5 &  &38.7$^{+.11}_{-.15}$   & $<$&38.9 	
& 38.2 		& 38.3	 	& 37.4$^{+.13}_{-.20}$ \\[0.5ex]
8   &  IRAS 01267--2157 & & 37.4$^{+.10}_{-.13}$ & & 37.7$^{+.13}_{-.18}$  
& &  37.7$^{+.06}_{-.07}$   & & 37.4$^{+.02}_{-.02}$	
& 37.3 		& 36.9	 	& 36.5$^{+.10}_{-.13}$ \\[0.5ex]
9   & RX J0134.2--4258 &
$<$&38.2	& &37.9$^{+.10}_{-.12}$	 & $<$&38.1 & $<$&38.3 	
& 37.9 		& 37.5$^{+.26}_{-.75}$	 & 34.0 \\[0.5ex]
10  & RX J0136.9--3510 &
$<$&38.3 	& $<$&38.1  & & 38.5$^{+.09}_{-.11}$ & $<$&38.7 	
& 37.4 		& 39.2	 	& 36.9$^{+.19}_{-.35}$ \\[0.5ex]
11  & RX J0148.3--2758 & $<$&37.5  	& &37.3$^{+.12}_{-.16}$	 & & 
38.0$^{+.05}_{-.05}$   & &37.9$^{+.10}_{-.13}$	
& 37.5 		& 37.7	 	& 37.1$^{+.06}_{-.07}$ \\[0.5ex] 
12  &  RX J0152.4--2319 & & 37.7$^{+.06}_{-.07}$ & & 37.5$^{+.08}_{-.09}$ 
& &  37.9$^{+.06}_{-.07}$   & &38.3$^{+.04}_{-.04}$  	
& 37.4 		& 37.1	 & 36.7$^{+.06}_{-.08}$ \\[0.5ex] 
13  & RX J0204.0--5104 &
$<$&37.3 	& $<$&37.4& $<$&37.5 & $<$&37.9 	
& 37.3 		& 37.6		& 36.8$^{+.11}_{-.15}$ \\[0.5ex] 
14  &  RX J0228.2--4057 & & 39.2$^{+.05}_{-.05}$ & & 39.3$^{+.02}_{-.02}$
 &  & 39.1$^{+.06}_{-.07}$   & & 39.2$^{+.10}_{-.12}$	
& 38.9 		& 38.7	 	& 38.0$^{+.10}_{-.12}$ \\[0.5ex] 
15  &  RX J0319.8--2627 & & 37.2$^{+.09}_{-.12}$  	
& & 37.4$^{+.05}_{-.05}$ & &  37.2$^{+.08}_{-.10}$   & $<$&37.5 
	& 37.0 		& 36.5	 	& 35.9$^{+.17}_{-.20}$ \\[0.5ex] 
16  & RX J0323.2--4931 & $<$&36.9 	& &
37.0$^{+.06}_{-.07}$	 &  &36.9$^{+.10}_{-.15}$   & $<$&37.3	
& 36.7 		& 37.0	 	& 36.4$^{+.05}_{-.06}$ \\[0.5ex] 
17  & ESO 301-G13 & & 37.2$^{+.06}_{-.07}$ 	& & 
37.2$^{+.03}_{-.03}$  & &  37.5$^{+.03}_{-.04}$   & $<$&37.2 
	& 36.9 		& 36.9	 	& 36.2$^{+.07}_{-.11}$ \\[0.5ex]
18  & VCV0331-373 & & 
36.9$^{+.06}_{-.07}$ 	& $<$&36.8 &  $<$&37.0   & $<$&37.2 	
& 36.6 		& 36.5	 	& 36.1$^{+.12}_{-.17}$ \\[0.5ex]
19  & RX J0349.1--4711 & &38.3$^{+.07}_{-.08}$ 	& & 
38.3$^{+.07}_{-.08}$  &  & 38.7$^{+.04}_{-.05}$   & $<$&38.7 
	& 37.9 		& 38.0	 	& 36.9$^{+.23}_{-.52}$ \\[0.5ex]
20  & Fairall 1116 & &35.8$^{+.11}_{-.15}$ & & 36.7$^{+.08}_{-.10}$	 
&  & 36.9$^{+.10}_{-.13}$   & & 37.3$^{+.07}_{-.08}$	
& 37.0 		& 37.4	 	& 36.5$^{+.09}_{-.11}$ \\[0.5ex]
21  & RX J0412.7--4712 & & 37.5$^{+.09}_{-.12}$ & & 37.8$^{+.03}_{-.04}$ 
& $<$&37.4 & & 37.7$^{+.12}_{-.17}$	
& 37.5 		& 37.2	 	& 36.7$^{+.16}_{-.25}$ \\[0.5ex]
22  & RX J0426.0--5112 & $<$&37.2 	& & 37.3$^{+.05}_{-.06}$	 
& & 37.5$^{+.05}_{-.05}$   & & 37.7$^{+.10}_{-.13}$	
& 37.9 		& 37.9	 	& 37.2$^{+.04}_{-.04}$ \\[0.5ex]
23  & Fairall 303 & $<$&36.4 	& &36.6$^{+.05}_{-.06}$	 
&  &36.6$^{+.06}_{-.07}$   & $<$&36.7 	
& 36.2 		& 36.2	 	& 35.8$^{+.09}_{-.11}$ \\[0.5ex]
24  & RX J0435.2--4615 & $<$&36.9 & &37.0$^{+.06}_{-.06}$	 
& $<$&36.9 & &37.6$^{+.05}_{-.06}$	
& 36.4 		& 36.4		& 35.7$^{+.14}_{-.21}$ \\[0.5ex]
25  & RX J0435.9--3636 & $<$&37.6 & &37.4$^{+.12}_{-.17}$ & $<$&37.8 & $<$&37.9
 	& 37.0 		& 37.0	 	& 36.7$^{+.11}_{-.14}$ \\[0.5ex]
26  & RX J0437.4--4711 & &
36.9$^{+.06}_{-.07}$ 	& $<$&36.6& $<$&36.6 & $<$&36.9 	
& 36.9 		& 36.6	 	& 36.0$^{+.08}_{-.09}$ \\[0.5ex]
27  & RX J0438.5--6148 & $<$&36.9 	& &36.9$^{+.04}_{-.04}$	 & & 
37.0$^{+.12}_{-.17}$   & & 37.8$^{+.03}_{-.04}$	
& 36.9 		& 36.8	 	& 36.3$^{+.04}_{-.04}$ \\[0.5ex]
28  & RX J0439.7--4540 & & 38.3$^{+.03}_{-.03}$	& & 38.1$^{+.06}_{-.07}$
	 & $<$&37.8 & & 38.3$^{+.11}_{-.15}$	
& 37.7	 	& 37.8$^{+.06}_{-.07}$	 	& 36.4 \\[0.5ex]
29  & RX J0454.7--4813 & 
$<$&38.3  	& &38.7$^{+.03}_{-.03}$	 & $<$&38.4 & $<$&38.8 	& 
37.6	 	& 37.7$^{+.13}_{-.18}$		& 36.8 \\[0.5ex]
30  & RX J1005.7+4332 & & 38.2$^{+.06}_{-.07}$ 	& & 38.4$^{+.02}_{-.02}$  
& & 38.2$^{+.06}_{-.06}$    & $<$&38.1 
	& 37.4	 	& 37.4	 	& 36.9$^{+.09}_{-.10}$ \\[0.5ex]
31  & CBS 126 & 
$<$&37.2 	& $<$&37.0& $<$&37.2 & $<$&37.3 	
& 37.1	 	& 37.0	 	& 36.6$^{+.05}_{-.06}$ \\[0.5ex]
32  & RX J1014.0+4619 &
$<$&38.3  	& $<$&38.1 & &  38.8$^{+.05}_{-.06}$    & $<$&38.7 	
& 37.9	 	& 37.4	 	& 36.8$^{+.23}_{-.36}$ \\[0.5ex]
33  & RX J1017.3+2914 & & 37.0$^{+.05}_{-.05}$ 	& & 37.0$^{+.04}_{-.04}$  
& & 37.1$^{+.05}_{-.06}$    & $<$&37.0 
	& 36.6	 	& 36.3	 	& 35.7$^{+.11}_{-.14}$ \\[0.5ex]
34  & Mkn 141 & & 37.4$^{+.02}_{-.02}$ 	& & 37.5$^{+.01}_{-.01}$
 & & 37.2$^{+.03}_{-.03}$    & & 37.5$^{+.03}_{-.04}$	
& 36.7	 	& 35.9	 	& 35.6$^{+.09}_{-.12}$ \\[0.5ex]
35  & Mkn 142 & & 36.8$^{+.09}_{-.11}$ 	& & 36.6$^{+.11}_{-.15}$ 
& & 36.6$^{+.12}_{-.17}$ & $<$&36.9 	
& 36.7	 	& 36.5	 	& 36.0$^{+.05}_{-.06}$ \\[0.5ex]
36  & RX J1050.9+5527 & & 38.8$^{+.04}_{-.04}$ 	& & 38.7$^{+.04}_{-.04}$ 
& & 38.6$^{+.05}_{-.06}$  & $<$& 39.1 	
& 38.0	 	& 37.7	 	& 37.1$^{+.16}_{-.25}$ \\[0.5ex]
37  & EXO 1055.3+6032 & $<$&37.6 	& & 37.7$^{+.06}_{-.07}$	 
& $<$& 37.7 & & 38.0$^{+.10}_{-.13}$	
& 37.1	 	& 36.8		& 36.3$^{+.14}_{-.20}$ \\[0.5ex]
38  & RX J1117.1+6522 & & 37.7$^{+.12}_{-.16}$ 	& & 37.7$^{+.07}_{-.08}$
	& & 38.0$^{+.05}_{-.06}$  &  & 38.5$^{+.04}_{-.05}$
	& 37.3	 	& 37.1		& 36.5$^{+.10}_{-.12}$ \\[0.5ex]
39   &  Ton 1388 & & 
---	& & --- 	 & & ---      & & --- 		
& 38.2	 	& 37.5	 	& 37.1$^{+.08}_{-.10}$ \\[0.5ex]
40   &  Mkn 734 & & 36.9$^{+.04}_{-.04}$ & & 36.8$^{+.03}_{-.03}$
 & & 36.9$^{+.08}_{-.10}$  & & 37.0$^{+.09}_{-.12}$	
	& 36.7	 	& 36.0	 	& 35.3$^{+.10}_{-.13}$ \\[0.5ex]
41   &  Z 1136.6+3412 & &
---	& & --- 	 & & --- 	    & & --- 		
& 36.1	 	& 36.1	 	& 35.6$^{+.09}_{-.11}$ \\[0.5ex]
42   &  CSO 109 & & 
---	& & --- 	 & & ---	    & & --- 		
& 36.5	 	& 36.3	 	& 35.9$^{+.09}_{-.12}$ \\[0.5ex]
43   & RX J1231.6+7044 & & 38.1$^{+.08}_{-.10}$	& & 37.9$^{+.08}_{-.10}$ 
& & 38.4$^{+.04}_{-.04}$     &$<$&39.1		
& 37.9	 	& 37.8		& 37.5$^{+.04}_{-.04}$ \\[0.5ex]
44   &IC 3599 & 
$<$&36.0	& & 35.8$^{+.07}_{-.08}$ &$<$&36.1   & $<$&36.3   	
& 35.6	 	& 37.0	 	& 35.7$^{+.05}_{-.06}$ \\[0.5ex]
45   & IRAS 12397+3333 & & 36.8$^{+.05}_{-.06}$	& & 37.1$^{+.02}_{-.02}$ 
& & 37.4$^{+.03}_{-.04}$     & & 37.0$^{+.10}_{-.13}$	 
	& 36.7	 	& 36.1	 	& 35.9$^{+.06}_{-.07}$ \\[0.5ex]
46   & RX J1312.9+2628 & 
$<$&36.8	& $<$&36.8    & & 37.0$^{+.09}_{-.11}$     & $<$&37.4 	
        & 36.6	 	& 36.2	 	& 35.9$^{+.09}_{-.11}$ \\
\noalign{\smallskip}\hline\noalign{}
\end{tabular}
\end{flushleft}
\end{table*}

\begin{table*}
\begin{flushleft}
\begin{tabular}{rlr@{}lr@{}lr@{}lr@{}llll}
\hline 
\noalign{\smallskip}\ \ 
\ \ No. & Name && $\nu L_{100\mu}$ && $\nu L_{60\mu}$ && $\nu L_{25\mu}$ && 
$\nu L_{12\mu}$ & $\nu L_{\rm V}$ & $\nu L_{250\rm eV}^{1)}$  & $\nu L_{\rm 1 
keV}^{1)}$  \\
\noalign{\smallskip}\hline\noalign{\smallskip}
47   & RX J1314.3+3429 & 
$<$&37.0	& $<$&37.0    &$<$&37.1   & $<$&37.6 	
        & 36.7	 	& 36.5	 	& 36.0$^{+.09}_{-.12}$ \\[0.5ex]
48   & RX J1355.2+5612 & & 37.5$^{+.04}_{-.05}$	& & 37.1$^{+.04}_{-.05}$    
& &37.9$^{+.04}_{-.05}$   &  & 37.2$^{+.04}_{-.05}$ 
& 37.0	 	& 37.1	 	& 36.5$^{+.08}_{-.10}$ \\[0.5ex]
49   & RX J1413.6+7029 & & 37.3$^{+.12}_{-.17}$	& & 37.6$^{+.05}_{-.06}$
	 &$<$& 37.2   & & 37.6$^{+.11}_{-.14}$	 	
& 36.9	 	& 37.1	 	& 36.8$^{+.05}_{-.05}$ \\[0.5ex]
50   & Mkn 684 & & 36.8$^{+.08}_{-.09}$     & & 36.9$^{+.05}_{-.05}$
 & & 36.9$^{+.05}_{-.05}$  & & 37.8$^{+.01}_{-.01}$	 
	& 36.9	 	& 36.0	 	& 35.8$^{+.08}_{-.09}$ \\[0.5ex]
51   & Mkn 478 & & 37.5$^{+.06}_{-.07}$	& & 37.6$^{+.05}_{-.05}$  
& & 37.2$^{+.08}_{-.10}$ & & 37.5$^{+.08}_{-.10}$	 	
& 37.4	 	& 37.5	 	& 36.8$^{+.04}_{-.04}$ \\[0.5ex]
52   & RX J1618.1+3619 & & 36.4$^{+.05}_{-.06}$	& & 35.9$^{+.11}_{-.15}$
	 &$<$&36.2   & $<$&36.7	 	
& 35.9	 	& 36.0	 	& 35.7$^{+.05}_{-.06}$ \\[0.5ex]
53   & RX J1646.4+3929 & &  37.4$^{+.08}_{-.10}$  & & 37.3$^{+.06}_{-.07}$
	 & & 37.5$^{+.06}_{-.07}$     & & 37.6$^{+.08}_{-.09}$	 	
& 36.6	 	& 36.7	 	& 36.3$^{+.09}_{-.11}$ \\[0.5ex]
54   & RX J2144.1--3949 & & 37.6$^{+.07}_{-.12}$ & & 37.5$^{+.10}_{-.12}$
	 &$<$&37.7   & $<$&38.0 	
 & 36.7	 	& 38.0	 	& 36.5$^{+.18}_{-.33}$ \\[0.5ex] 
55   & RX J2154.1--4414 & $<$&38.6  & $<$&38.3   
& & 38.9$^{+.04}_{-.05}$     & $<$&38.7 	  
      & 38.3	 	& 38.3	 	& 37.3$^{+.18}_{-.31}$ \\[0.5ex]
56   & RX J2213.0--1710 & $<$&37.5  & & 37.6$^{+.08}_{-.10}$	 
& & 38.0$^{+.08}_{-.10}$     & $<$& 38.2 	        
& 37.0	 	& 37.5	 	& 36.6$^{+.16}_{-.24}$ \\[0.5ex]
57   & RX J2216.9--4451 & $<$&37.8	& $<$&37.4    
& &  38.2$^{+.04}_{-.05}$     & & 38.3$^{+.03}_{-.03}$	 	
& 37.4	 	& 37.8	 	& 36.9$^{+.09}_{-.11}$ \\[0.5ex]
58   & RX J2217.9--5941 & & 38.7$^{+.03}_{-.03}$ & & 38.5$^{+.01}_{-.01}$
	 & & 38.2$^{+.04}_{-.05}$	    & $<$&38.3	 	
& 37.4	 	& 37.8	 	& 36.8$^{+.11}_{-.14}$ \\[0.5ex]
59   & RX J2221.8--2713 & & 37.9$^{+.11}_{-.14}$  & & 37.7$^{+.10}_{-.12}$
	 & $<$&38.0  & $<$&38.2 	       
 & 37.0	 	& 37.1	 	& 36.5$^{+.21}_{-.31}$ \\[0.5ex]
60   & RX J2232.7--4134 & $<$&37.1  & & 36.9$^{+.11}_{-.15}$	
 &$<$&37.3   & & 37.8$^{+.05}_{-.05}$	 	
& 36.5	 	& 36.2	 	& 35.7$^{+.20}_{-.29}$ \\[0.5ex]
61   & RX J2241.8--4405 & 
$<$&39.0	& $<$&38.7    & $<$&38.8  & & 39.4$^{+.08}_{-.09}$
	 & 38.8	 	& 38.7	 	& 37.8$^{+.20}_{-.37}$ \\[0.5ex]
62   & RX J2242.6--3845 & $<$&38.1  & $<$&37.9     
& & 38.2$^{+.12}_{-.16}$     & $<$&38.4 	        
& 37.5	 	& 37.6	 	& 36.9$^{+.14}_{-.21}$ \\[0.5ex]
63   & RX J2245.3--4652 & & 38.2$^{+.01}_{-.01}$ & & 38.4$^{+.03}_{-.03}$
	 & & 38.2$^{+.10}_{-.12}$     & & 38.7$^{+.06}_{-.07}$	 	
& 38.3	 	& 38.2	 	& 37.3$^{+.09}_{-.10}$ \\[0.5ex]
64   & RX J2248.6--5109 & &  37.6$^{+.07}_{-.09}$ & & 37.7$^{+.03}_{-.03}$
	 & & 37.4$^{+.11}_{-.15}$     & $<$&37.7 	        
& 37.4	 	& 37.5	 	& 36.9$^{+.06}_{-.07}$ \\[0.5ex]
65   & MS2254.9--3712 & & 36.8$^{+.03}_{-.04}$     & & 36.9$^{+.03}_{-.03}$
	 & & 36.8$^{+.09}_{-.12}$	    & $<$&37.1 	        
& 36.7	 	& 36.4	 	& 35.9$^{+.08}_{-.10}$ \\[0.5ex]
66   & RX J2258.7--2609 &
$<$&37.1  & $<$&37.0& $<$&37.4  & $<$&37.5 	        
& 36.8	 	& 36.7	 	& 36.5$^{+.12}_{-.17}$ \\[0.5ex]
67   & RX J2301.8--5508 & & 38.0$^{+.06}_{-.07}$  & & 38.0$^{+.02}_{-.02}$
	 & & 38.0$^{+.04}_{-.04}$	    & & 38.1$^{+.08}_{-.10}$
& 37.7	 	& 37.4	 	& 36.8$^{+.09}_{-.11}$ \\[0.5ex]
68   & RX J2303.9--5517 &  $<$&37.1	& $<$&36.7    
&&  37.3$^{+.09}_{-.11}$    & $<$&37.5      	
& 36.4	 	& 36.4	 	& 36.2$^{+.10}_{-.13}$ \\[0.5ex]
69   & RX J2304.6--3501 & $<$&36.6  & & 36.9$^{+.03}_{-.03}$	
 & & 36.8$^{+.05}_{-.06}$     & $<$&36.7 	      
  & 36.1	 	& 36.0	 	& 35.7$^{+.18}_{-.30}$ \\[0.5ex]
70   & RX J2304.6--5128 & & 38.1$^{+.02}_{-.02}$ & & 38.1$^{+.02}_{-.02}$
	 & & 37.6$^{+.07}_{-.09}$     & $<$&37.6 	       
 & 36.7	 	& 36.7	 	& 36.2$^{+.14}_{-.20}$ \\[0.5ex]
71   & RX J2317.8--4422 & & 37.7$^{+.10}_{-.13}$ & & 37.7$^{+.07}_{-.09}$
	 & & 38.0$^{+.05}_{-.05}$     & $<$&37.8 	    
    & 37.0	 	& 37.4	 	& 36.4$^{+.18}_{-.32}$ \\[0.5ex]
72   & RX J2325.2--3236 & $<$&38.3	& $<$&38.1      
& & 38.4$^{+.08}_{-.09}$     & $<$&38.8 	  
      & 37.4	 	& 37.6	 	& 36.9$^{+.22}_{-.47}$ \\[0.5ex]
73   & RX J2340.6--5329 &  $<$&38.6	& & 38.1$^{+.11}_{-.14}$	
 & & 38.7$^{+.09}_{-.11}$     & $<$&38.7 	
        & 37.5	 	& 37.9	 	& 37.2$^{+.27}_{-.90}$ \\[0.5ex]
74   & MS2340.9--1511 & & 37.9$^{+.08}_{-.09}$     & $<$&37.5   
   &$<$&37.8   & & 38.4$^{+.05}_{-.06}$	 	
& 37.6	 	& 38.0	 	& 36.8$^{+.14}_{-.21}$ \\[0.5ex]
75   & RX J2349.1--3311 & & 
---      & & ---    & & ---	    & & --- 		
& 37.1	 	& 36.9	 	& 36.6$^{+.11}_{-.15}$ \\[0.5ex]
76   & RX J2349.3--3125 & & 38.0$^{+.09}_{-.12}$    & &37.8$^{+.05}_{-.06}$
	 &$<$&37.7   & & 38.3$^{+.05}_{-.05}$	 	
& 37.2	 	& 37.1	 	& 36.7$^{+.11}_{-.15}$  \\
\noalign{\smallskip}\hline\noalign{\smallskip}
\end{tabular}

\noindent {\footnotesize $^{1)}$
Errors based on the statistical uncertainties in the spectral flux at the
reference energy of the power-law fits, uncertainties in \ax~ not included 
}
\end{flushleft}
\end{table*}
\normalsize

\footnotesize
\begin{table*}[ht]
\caption[Mean luminosities and spectral indices] {\label{mean} Means and
standard errors of the means, median values, and 90\% ranges. Luminosities are
at rest-frame wavelengths or energies, respectively. Only the 71 soft X-ray AGN 
for which IRAS data were available were used. See text for treatment of 
upper limits in Table~\ref{lum-list}.} 
\begin{center}
\begin{tabular}{lrrcrrc}
\noalign{\smallskip} \hline \noalign{\smallskip}
& \multicolumn{3}{c}{soft X-ray AGN} &  \multicolumn{3}{c}{hard X-ray AGN} \\[0.5ex]
& mean & median & 90\% range & mean & median & 90\% range \\
\noalign{\smallskip}\hline\noalign{\smallskip}
log $\nu L_{100\mu}$ & 37.11\pl 0.12 &  &  & 
36.89\pl 0.14  & 36.85 & 36.2-37.8 \\
\noalign{\smallskip}
log $\nu L_{60\mu}$ & 37.31\pl0.10  &  &  &
36.98\pl 0.13 & 36.86 & 36.2-37.8  \\
\noalign{\smallskip}
log $\nu L_{25\mu}$ & 37.40\pl0.10 &  &  &
 37.10\pl 0.14 & 36.99 & 36.2-37.8  \\
\noalign{\smallskip}
log $\nu L_{12\mu}$ & 37.23\pl0.12 &  &  & 
 37.16\pl0.12 & 37.19 & 36.4-37.8 \\
\noalign{\smallskip}
log $\nu L_{\rm V}$ & 37.19$\pm$0.07  & 37.12 & 36.0-38.2 &
37.10$\pm$0.12  & 36.98 & 36.4-37.8 \\
\noalign{\smallskip}
log $\nu L_{\rm 250eV}$ & 37.17$\pm$0.08  & 37.10 & 36.0-38.2 & 
36.68$\pm$0.16 & 36.77 & 35.6-37.4 \\
\noalign{\smallskip}
log $\nu L_{\rm 1keV}$ & 36.49$\pm$0.08  & 36.54 & 35.6-37.4 &
36.44$\pm$0.15 & 36.51 & 35.6-37.0 \\
\noalign{\medskip}
\ax & 2.11$\pm$0.10 & 1.96 & 1.3-2.7 & 1.35$\pm$0.06 & 1.40 & 1.0-1.7  \\
\noalign{\smallskip}
$\alpha_{\rm opt}$ & 0.96$\pm$0.08  & 1.03 & 0.0-1.9 & 
1.37$\pm$0.20 & 1.31 & 0.3-2.1 \\
\noalign{\smallskip}
$\alpha_{\rm ox-soft}$ & 1.04$\pm$0.02  & 1.05 & 0.7-1.2 &
1.21$\pm$0.05 & 1.22 & 0.9-1.5 \\
\noalign{\smallskip}
$\alpha_{\rm ox-hard}$ & 1.26$\pm$0.02 & 1.25 & 1.1-1.4 & 
1.25$\pm$0.04 & 1.26 & 1.0-1.4 \\
\noalign{\smallskip}
$\alpha_{\rm 60\mu-opt}$ & 1.11\pl0.03  &  &  & 
0.94\pl0.04 & 0.95 & 0.6-1.3 \\
\noalign{\smallskip}
$\alpha_{\rm 60\mu-soft}$ & 1.05\pl0.02  & &  &
1.08\pl0.03 & 1.08 & 0.9-1.2 \\
\noalign{\smallskip}
$\alpha_{\rm 60\mu-hard}$ & 1.19\pl0.02  &  &  &
1.12\pl0.03 & 1.12 & 1.0-1.3 \\
\noalign{\smallskip}
$\alpha_{\rm 12\mu-opt}$ & 1.23\pl0.05  &  &  &
1.05\pl0.05 & 1.08 & 0.7-1.3 \\
\noalign{\smallskip}
$\alpha_{\rm 12\mu-soft}$ & 1.08\pl0.03  &  &  & 
1.14\pl0.03 & 1.15 & 0.9-1.4 \\
\noalign{\smallskip}
$\alpha_{\rm 12\mu-hard}$ & 1.25\pl0.02  & &  &
1.18\pl0.02 & 1.19 & 1.1-1.3 \\
\noalign{\medskip}
z & 0.144$\pm$0.013 & 0.113 & 0.03-0.30 & 0.050\pl0.011 & 0.033 & 0.01-0.10 \\
\noalign{\smallskip}\hline\noalign{\smallskip}
\end{tabular}
\end{center}
\end{table*}
\normalsize

\begin{figure}[t]
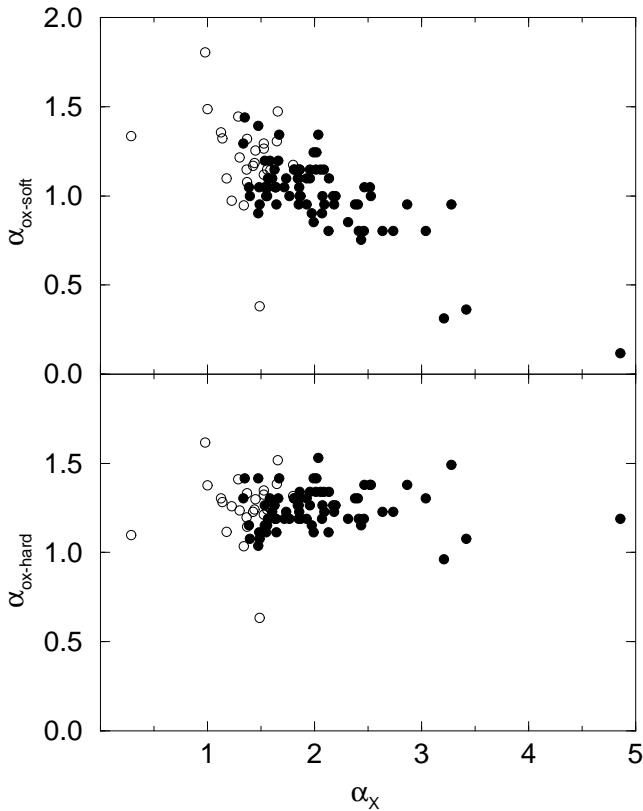

\clipfig{alpha_opt-x}{87}{40}{8}{164}{165}
\caption[ ]{\label{alpha_opt-x}
Optical to X-ray spectral indices \aoxs~ and \aoxh~vs. X-ray slope \ax. 
WPVS007, RX J0134--42, and RX J0136--35 
are off the plot (\ax=8.0, $\alpha_{\rm ox-soft}$=1.2,  
$\alpha_{\rm ox-hard}$=2.7; 6.7, 1.2, 2.5; 4.9, 0.6, 1.2, respectively).
Solid circles denote AGN from the soft and open circles from
the hard comparison sample.
}
\end{figure}

The observation that \aoxh~ remains constant independent of \ax, whereas \aoxs~
becomes flatter for the steeper X-ray spectra (Fig.~\ref{alpha_opt-x}) is 
suggestive of an underlying spectrum which connects optical and 1~keV fluxes.
The slope of this steep-spectrum continuum component is $1.3$ for soft {\it and}
hard X-ray selected samples. The high mean 250~eV flux would then represent an
emission component superimposed on the nuclear steep-spectrum component 
(i.e.  a
soft excess). It appears, therefore, that a disk-like BBB emission component
competing with a nuclear steep-spectrum component becomes increasingly more
apparent at optical and soft X-ray wavelengths as the luminosity increases.

\section{Discussion}

Normal Sy1 galaxies typically show a Big Blue Bump which is most pronounced in
the UV (Walter et al.  1994). In the optical and soft X-ray ranges, the BBB
competes with an underlying continuum which appears above keV energies with the
canonical hard X-ray slope of 0.7 ($\sim$1.0 after subtraction of a reflection
component) and which has a steep spectrum in the optical. Generally, a slope of
$\sim$~1.3 connects the optical and keV energy ranges.  Since the optical slopes
are often much steeper than 1.3, the optical and X-ray continua are not part of
the same spectral component, but seem to be energetically linked. In the
infrared, part of the BBB emission seems to be reprocessed by a circum-nuclear
dust torus (Krolik 1996). 

\subsection{\label{dis_soft} Soft X-ray excess or lack of hard X-rays?}

\footnotesize
\begin{table*}
\caption
{\label{rank} Spearman rank-order 
correlation
coefficients $r_{\rm s}$ (below diagonal)
and significance levels $t$ (above diagonal) for $N=71$. Negative values
denote an anti-correlation and  positive values a correlation.
}
\begin{flushleft}
\begin{tabular}{lcccccccc}
\noalign{\smallskip} \hline \noalign{\smallskip}
& log \vlir & log \vlv & log \vlsoft & \ax & \aopt & \aoxsoft &
\dnh & rank(\ax)$-$rank(\dnh) 
\\
\noalign{\smallskip}\hline\noalign{\smallskip}
log \vlir   & 1     & 8.6   & 8.6   & 3.1   & $-3.2$   & $-0.9$   &
 $-$1.4 & 2.9 \\
\noalign{\smallskip}
log \vlv    & 0.72  & 1     & 14.0  & 3.8   & $-5.3$   & $-0.2$  
& $-$1.6 & 2.6 \\
\noalign{\smallskip}
log \vlsoft & 0.72  & 0.86  & 1     & 6.7   & $-5.3$   & $-3.9$  
& $-$2.0 & 4.3 \\
\noalign{\smallskip}
\ax         & 0.35  & 0.42  & 0.63  & 1     & $-3.9$   & $-4.7$   
& $-2.2$ & 11.4 \\ \noalign{\smallskip}
\aopt       & $-0.36$ & $-0.54$ & $-0.54$ & $-0.42$ & 1     & 0.6   
& 3.1 & $-$4.6 \\
\noalign{\smallskip}
\aoxsoft    & $-0.11$ & $0.02$ & $-0.43$ & $-0.49$ & 0.07  & 1     
& 1.0 & $-$3.0 \\
 \noalign{\smallskip}
\dnh & $-0.17$ & $-0.19$ & $-0.23$ & $-0.26$ & 0.35 & 0.12 &  
1 & $-$11.4 \\ \noalign{\smallskip}
rank(\ax)$-$rank(\dnh) & 
0.33 & 0.30 & 0.46 & 0.81 & $-$0.48 & $-$0.34 & $-$0.81 & 1 \\
\noalign{\smallskip}\hline\noalign{\smallskip}
\end{tabular}
\end{flushleft}
\end{table*}
\normalsize

In general, the steep X-ray spectra of soft X-ray selected AGN could indicate
(i) an excess of soft X-rays over the canonical (flat) hard X-ray continuum 
(i.e. an extension of the BBB into the X-ray range) or
(ii) a systematic lack of emission around 1 keV and possibly above.  Several
authors favored case (i) (e.g.  C\'ordova et al.  1992, Puchnarewicz et al.
1992, Walter \& Fink 1993, Boller et al.  1996), while Laor et al. (1994) argued
for case (ii) based on their ROSAT PSPC observations of PG quasars.  Laor et al.
 suggested that ``a steep \ax~is associated with a weak hard X-ray component,
relative to the near-IR (60$\mu$m) and optical emission, rather than a soft
X-ray excess''. Warm absorbers are also in line with case (ii) 
(see Sect.~5.3). 

The new soft X-ray selected AGN seem to follow case (i), since the 250~eV
luminosities are clearly enhanced relative to the optical and hard X-ray
luminosities compared with the hard X-ray selected AGN, whereas the
optical-to-1-keV spectral indices are the same (1.3)  (Fig.~3, Table~2). Pointed
ROSAT observations with higher statistical significance corroborate the
existence of a separate hard X-ray component at least in some of our objects.
Whether an underlying hard X-ray continuum indeed exists in general for soft
X-ray AGN remains open\footnote{The soft X-ray selected Seyfert galaxy RX
J1034+39 (Z 1031.7+3954), e.g., has a steep ASCA spectrum which continues up to
at least 10 keV without flattening (Pounds et al. 1995).}. 

\subsection{\label{dnh} \dnh~ as a Measure of the Soft X-ray Excess}

We have seen that the intrinsic absorption is low in the soft X-ray AGN.
However, the picture becomes a bit more complicated on a second view. The
quantities \nh~ as well as \ax~ were measured from single power-law fits. While
this may be an acceptable first-order approximation, many AGN possess a soft
excess over a flatter hard X-ray spectrum (see e.g. Leighly et al. 1996,
Reynolds 1997, Walter et  al. 1994). The result is a negative curvature over the
ROSAT spectral range. The excess is then measured both by a larger \ax~ and
$N_{\rm H,fit}<N_{\rm H,gal}$ (negative \dnh). The slope \ax~ is weighted
towards higher energies, while \dnh~ is most affected by the excess at the
lowest energies. Therefore \dnh~ is in principle affected both by intrinsic
absorption and an excess of soft X-ray emission. Puchnarewicz et al. (1995b)
discussed the case of the soft AGN RX J2248--51 with $N_{\rm
H,fit}<N_{\rm H,gal}$. They showed this to be a result of a steepening of the
X-ray spectrum towards lower energies rather than a
`hole' in the interstellar gas at the position of this source. 
RX J2248--51 is also in contained in our sample. 

We find marginal evidence for a relation between \dnh~and \aopt~ in the way
expected if the X-ray spectra increasingly steepen towards ~0.2 keV in the
sources with the flattest optical spectra (Table~4).  The significance of the
correlation is only $t=3.1$, but it is increased for the anti-correlation
between the combined rank [Rank(\ax) $-$ Rank(\dnh)] and Rank(\aopt), which
yields $t=-4.6$.  If \dnh~ were unrelated to the soft excess, one would have
expected a lower significance for the combined ranks than for the individual
ones. Of course, it is difficult to find significant deviations from single
power law spectra on the basis of RASS PSPC spectra alone which only contain a
few hundred photons typically.  Further analysis of higher signal-to-noise
pointed PSPC spectra, available for some of the AGN in our sample, is required
to investigate this effect in more detail. 

\subsection{Origin of the optical and X-ray emission components}

One of our main findings is that the new soft X-ray selected AGN are Sy1
galaxies with an enhanced BBB emission component relative to the underlying
continuum making the optical continuum blue and the soft X-ray spectrum steep
(Fig.~\ref{sed_example}). Puchnarewicz et al.~(1996) recently reported a similar
\ax--\aopt~relation in the RIXOS AGN sample which supports our result. The
strong correlation between \ax~ and $\nu F_{\rm 1375\AA}/\nu F_{\rm 2keV}$ for
Sy 1 nuclei in general (Walter \& Fink 1993) is another way of presenting the
same relation between the BBB and soft X-rays, and our results support their
interpretation of this soft X-ray excess.  Although less significant, it seems
that the IR emission is also enhanced which would indeed be expected if
re-processing of the BBB emission by dust is important. The extreme blueness of
some of the optical spectra ($\alpha_{\rm opt}$ approaches $-1/3$) which has also
been found in other AGN (Richstone \& Schmidt 1980, Sun \& Malkan 1989), is in
agreement with emission from an optically thick, geometrically thin accretion
disk. 

The enormous width of the BBB requires additional comptonization of the
optically thick disk spectrum by a hot corona (Matt et al.  1993) or
near-Eddington accretion (Ross et al.  1992) both of which lead to copious soft
X-ray production. Whether the large dispersion in the observed soft X-ray slopes
represents a true dispersion in the emission process or is rather the
consequence of the superposition of a (possibly Wien-shaped) ultrasoft X-ray
component with a separate flatter hard X-ray component remains an open issue in
the absence of hard X-ray data. 

The ubiquitous $\alpha_{\rm ox-hard} \simeq 1.3$ argues for a common origin of
the steep-spectrum optical emission component and the hard X-ray
component\footnote{ A correlation $L_{\rm opt}\propto L_{\rm x}$ seems to hold
for AGN in general (La Franca et al.~1995).}. In some AGN the optical spectrum
is steeper than the optical-to-X-ray spectrum which is inconsistent with a
single non-thermal continuum unless there is reddening in these sources.  We
have no evidence, however, for systematic reddening effects from our
low-resolution identification spectra. 

An optical continuum component which does not connect as a power law with the
hard X-ray part of the spectrum, but is still energetically coupled to the hard
X-ray emission component, could result from thermal re-processing of the hard
X-rays  by cold matter. The cold matter intercepting hard X-rays  may be located
in the outer accretion disk (Collin-Souffrin 1987).  In order to re-process a
large fraction of the hard X-ray power, the solid angle subtended by the disk
with respect to the hard X-ray source must be large.  A plausible geometry would
be that the hard X-ray source is located {\it above} a geometrically thin disk
at a height comparable to the optical emission radii of the disk $\sim 10^2-10^4
R_{\rm G}$ where $R_{\rm G}=2GM/c^2$ denotes the Schwarzschild radius.  This
geometry is in rough agreement with the one found by Clavel et al.~(1992) from
multi-wavelength variability studies of the hard X-ray AGN NGC~5548 from which a
height of $\sim 100 R_{\rm G}$ was inferred. 

The predominance of the BBB appears to increase with the observed soft X-ray
luminosity which, in turn, depends on the accretion rate and on the viewing
angle. A variation of black hole mass keeping \dMM$\propto \dot{M}/\dot{M}_{\rm
Edd} $ constant does not seem to be responsible for the variable bump strength. 
With increasing luminosity, the inner disk temperature would decrease according
to $T_{*}\propto M^{-{1\over 4}}$ and would, therefore, fail to explain the
observed soft X-ray excess unless the effect is compensated by comptonization
(e.g., Czerny \& Elvis 1987).  Also, the variability time scales would be
expected to increase with luminosity which does not seem to be the case.  This
effect would be quite dramatic, since the most luminous AGN in our sample would
require black hole masses a factor of $\sim$100 higher than in the weakest
sources. 

It is more likely that the effect is due to an increasing $\dot{M}/\dot{M}_{\rm
Edd}$ and that the  most luminous soft X-ray selected AGN are close to Eddington
accretion. The soft X-ray excess would be related to the inner disk temperature
increasing as $T_{*}\propto  \dot {M}^{{1\over 4}}$. There would be a striking
analogy between AGN and Galactic Black Hole candidates such as Cyg X-1 (Pounds
et al.~1995) which display `high' states with a thermally dominated UV/soft
X-ray spectrum at a near-Eddington accretion rate and `low' states in which the
hard X-rays are much more pronounced and the accretion rate is reduced.
RX~J0134-42 has indeed shown a turn-on of the hard X-ray component with an
associated decrease in the strength of the BBB (Mannheim et al.~1996). Also the
transient sources WPVS007 (Grupe et al. 1995b) and IC3599 (Brandt et al.~1995, 
Grupe et al.~1995a) may fit into this picture. In fact, any transient Eddington
high state produced by a $\ga 10^6M_\odot$ black hole would be located to the
right of log\,\lx $\simeq$\,log\,\lsoft$\approx37$ in Fig.~6 and could be
responsible for the large dispersion in \ax. 

Further support for the high state scenario may come from the emission line
properties of the soft AGN (which we will describe in more detail in Paper II).
There exists a relation between \ax~and FW(\hb) in the sense that steep X-ray
spectra are observed preferentially in systems with narrow permitted lines
(Boller et al. 1996, Paper II).  Boller et al.~argue that narrow-line Sy1s have
moderate black hole masses in the range $\sim 10^6-10^7M_\odot$ but accretion
rates \dMM~ higher than in normal Sy1s. The enhanced photoionizing flux destroys
the emission line clouds closest to the central black hole. 

Alternative explanations of the \ax--FW(\hb) relation do not seem to be
consistent with our results. If the emission line clouds originate in a torus
geometry, the smallest line widths would  be expected for a face-on orientation,
i.e. the relation would be a mere orientation effect.  There would be no
absorbing cold gas and therefore steep X-ray spectra and blue optical continua.
With increasing viewing angle, one would expect a hardening of the X-ray spectra
owing to absorption by the gaseous torus. However,  intrinsic absorption by cold
gas is not responsible for the slopes of the X-ray spectra. Absorption by warm
dust reddening the optical spectra and steepening the X-ray spectra  would have
the opposite effect of what is needed. The soft AGN could also correspond to
relativistic accretion disks seen near the disk plane where the  X-ray (but not
the optical) flux  is Doppler enhanced.  Unless the emission line gas is moving
preferentially perpendicular to the disk (e.g., a disk wind), there is, however,
no reason to expect smaller line widths in these AGN. While it is clear that the
soft X-rays are not viewed through a dusty torus with large column density, it
is not possible to further constrain the geometry and inclination based on the
continuum properties alone. 

It is undisputed that the large range of luminosities in AGN generally requires
a range in black hole masses, but it appears plausible that the peculiar
properties of the soft X-ray selected Seyferts are due to a temporarily
increased $\dot{M}$ for a rather narrow range in $M$. 

\subsection{\label{dis_warm} Warm absorbers} 

A primary X-ray spectrum could appear steeper in the soft X-ray band than the
original spectrum if it is absorbed by partially ionized matter at intermediate
energies between $\sim$0.5 and 3 keV, which becomes optically thin for soft
X-rays below 0.5~keV.  Netzer (1993) has shown that the deep edges are
practically washed out in realistic models for the geometry of a warm absorber
taking into account emission lines and reflection.  Several observations, such
as the ROSAT observations of IRAS 13349+2438 (Brandt et al.~1996) or the 3C351
observations of Fiore et al. (1993), have led to claims that this warm
absorption could be important in mimicking a soft excess (see also Komossa \&
Fink 1997). With ASCA, evidence for warm absorbers in AGN has been found in the
form of the OVI, OVII, and OVIII absorption edges (e.g. George et al.~1995,
Mihara et al.~1995, Otani et al.~1996) without, however, obliterating the need
for primary X-ray spectra steeper than the canonical hard X-ray spectra.  The
spectral steepening is maximized if much of the absorber is optically thin to
radiation below the K- and L-edges of the more abundant elements, causing the
spectrum to be selectively depressed around 1\,keV. Although warm absorbers are
present in some of the soft X-ray emitting AGN (see Brandt et al. 1996, we will
also discuss this issue on the polarized AGN of our sample in two separate
papers, Grupe et al. 1997 and Wills et al. in prep.), there is no evidence on
the average for an association of steep spectra and low 1\,keV fluxes
($\alpha_{\rm ox-hard}=$~const.). Therefore, warm absorbers cannot be the
primary cause for the very soft X-ray spectra of our soft sample. 

\section{Conclusion}

The interpretation of the BBB emission component in AGN in terms of an accretion
disk has generally suffered from the presence of additional optical and hard
X-ray emission components radiating comparable amounts of energy.  Our data
demonstrate that soft X-ray selection provides a powerful means for discovering
AGN with almost bare BBB emission characterized by blue optical spectra with a
slope approaching $\alpha_{\rm opt}\approx -1/3$ and very soft X-ray spectra. We
speculate that these very soft X-ray AGN correspond to $10^6-10^7M_\odot$ black
holes in a high state with near-Eddington accretion. The phenomenon is likely to
be of a transient nature, basically similar to what has been observed in IC3599,
WPVS007, and RXJ0134-42, although on longer time scales. Further studies in the
UV and EUV bands are required to measure the shape and variability of the BBB in
the wavelength regime where most of the bolometric luminosity is apparently
released.  It is also important to clarify the role of absorption by warm dust,
the nature of the underlying continuum, and its connection to the BBB by
additional hard X-ray and near-infrared observations.

\acknowledgements{We thank B. Wills and the anonymous referee for useful
comments. We are also grateful to assistance by the ROSAT team at the MPE. We
used the NASA/IPAC Extragalactic Database which is operated by the Jet
Propulsion Laboratory, Caltech, under contract with the National Aeronautics and
Space Administration, and IRAS data provided by the Infrared Processing and
Analysis Center, Caltech. This research was initially supported by the DARA
grant 50 OR 92 10 and later on by the Bundesanstalt f\"ur Arbeit. }

\end{document}